\documentclass[10pt]{article}

\usepackage{amsmath}
\usepackage{amssymb}
\usepackage{textcomp}
\usepackage{graphicx}

\usepackage{cite}

\usepackage{color} 

\usepackage{setspace} 
\usepackage[labelfont=bf,labelsep=period,justification=raggedright]{caption}

\makeatletter
\renewcommand{\@biblabel}[1]{\quad#1.}
\makeatother

\date{}

\begin{document}

\begin{flushleft}
{\Large
\textbf{Modulation of calmodulin lobes by different targets: an allosteric model with hemiconcerted conformational transitions}
}
\\
Massimo Lai$^{1\ast}$, 
Denis Brun$^{2,3}$,
Stuart J. Edelstein$^{1}$ and 
Nicolas Le Nov\`ere$^{1,2}$
\\
\bf{1} Babraham Institute, Cambridge, United Kingdom \\
\bf{2} EMBL-EBI, Hinxton, United Kingdom \\
\bf{3} currently at Amadeus IT Group, Sophia Antipolis, France\\
$\ast$ E-mail: massimo.lai@babraham.ac.uk
\end{flushleft}

\section*{Abstract}
Calmodulin is a calcium-binding protein ubiquitous in eukaryotic cells, involved in numerous calcium-regulated biological phenomena, such as synaptic plasticity, muscle contraction, cell cycle, and circadian rhythms. It exibits a characteristic dumbell shape, with two globular domains (N- and C-terminal lobe) joined by a linker region. Each lobe can take alternative conformations, affected by the binding of calcium and target proteins.
Calmodulin displays considerable functional flexibility due to its capability to bind different targets, often in a tissue-specific fashion. In various specific physiological environments (e.g. skeletal muscle, neuron dendritic spines) several targets compete for the same calmodulin pool, regulating its availability and affinity for calcium. In this work, we sought to understand the general principles underlying calmodulin modulation by different target proteins, and to account for simultaneous effects of multiple competing targets, thus enabling a more realistic simulation of calmodulin-dependent pathways. We built a mechanistic allosteric model of calmodulin, based on an hemiconcerted framework: each calmodulin lobe can exist in two conformations in thermodynamic equilibrium, with different affinities for calcium and different affinities for each target. Each lobe was allowed to switch conformation on its own. The model was parameterised and validated against experimental data from the literature. In spite of its simplicity, a two-state allosteric model was able to satisfactorily represent several sets of experiments, in particular the binding of calcium on intact and truncated calmodulin and the effect of different skMLCK peptides on calmodulin's saturation curve. The model can also be readily extended to include multiple targets. We show that some targets stabilise the low calcium affinity T state while others stabilise the high affinity R state. Most of the effects produced by calmodulin targets can be explained as modulation of a pre-existing dynamic equilibrium between different conformations of calmodulin's lobes, in agreement with linkage theory and MWC-type models.

\section*{Author Summary}
Calmodulin, the ubiquitous calcium-activated second messenger in eukaryotes,
is an extremely versatile molecule involved in many biological processes:
muscular contraction, synaptic plasticity, circadian rhythm, and cell cycle, among others. The protein is structurally organised into two globular lobes, joined by a flexible linker. 
Calcium modulates calmodulin activity by favoring a conformational transition 
of each lobe from a closed conformation to an open conformation. Most targets 
have a strong preference for one conformation over the other, and depending on  
the free calcium concentration in a cell, particular sets of targets will preferentially interact with calmodulin. In turn, targets can increase or decrease 
the calcium affinity of the calmodulin molecules to which they bind. 
Interestingly, experiments with the tryptic fragments showed that most targets have a much 
lower affinity for the N-lobe than for the C-lobe. Hence, the latter predominates 
in the formation of most calmodulin-target complexes.
We showed that a relatively simple allosteric mechanism, based the classic MWC model, 
can capture the observed modulation of both the isolated C-lobe, and intact calmodulin, 
by individual targets. 
Moreover, our model can be naturally extended to study how the calcium affinity of a
single pool of calmodulin is modulated by a mixture of competing targets \emph{in vivo}. 

\section*{Introduction}
Calmodulin is a ubiquitous calcium-binding protein involved in many cellular
processes, from synaptic plasticity to muscular contraction,
cell cycle regulation, and circadian rhythms. Structurally, it is organized in
two highly homologous globular domains joined by a flexible linker
\cite{Faga2003}. 
Each domain contains two calcium-binding EF-hands that can undergo a transition
between closed and open conformations, the latter favored
by calcium binding \cite{Zhang1995,Grabarek2005}. The transition to the open
state results in exposure of hydrophobic residues able to interact with numerous
binding partners \cite{Nelson1998}.
However, some targets interact preferably with the closed form
\cite{Rhoads1997}.

The shift of calmodulin's calcium saturation curve in the presence 
of targets was studied experimentally by several groups
\cite{Peersen1997,Bayley1996,Gaertner2004,Evans2009,Feldkamp2011}.
Targets that markedly increase calmodulin's calcium affinity include the calcium-calmodulin dependent kinase II (CaMKII), protein phosphatase 2B (PP2B) and skeletal muscle myosin light chain kinase (skMLCK). It was shown that, in general, targets do not bind calmodulin exclusively in either calcium-saturated or calcium-free forms, but rather bind to both forms with different affinities. Numerous biologically relevant targets exhibit this behavior, such as skMLCK, CaMKII, and the NaV1.2 sodium channel \cite{Bayley1996,Feldkamp2011,Kim2005,Theoharis2008,O'Donnell2011}.
The binding domains that preferably interact with calcium-free calmodulin are called IQ motifs, a family of 14-residue sequences named after the two most frequent initial amino acids (usually isoleucine, followed by a highly conserved glutamate) \cite{Rhoads1997}.

Calmodulin's properties, such as cooperativity and affinity modulation 
by binding targets, can be explained by an MWC allosteric model.
The first model of allosteric transitions was introduced in the seminal paper 
by Monod, Wyman and Changeux in 1965, which dealt with multimeric proteins
whose subunits could undergo concerted conformational transitions, and also be
modulated by target binding either to the R state, or the T state, in an
exclusive fashion \cite{Monod1965}.The model was then further extended by Rubin and Changeux to describe molecules that could be modulated by binding partners capable of binding both
conformational states, but with different affinites \cite{Rubin1966}.
The hypothesis of the existence of distinct calmodulin conformations in thermodynamic equilibrium, which is crucial to an allosteric model, is supported by 
evidence of conformational dynamics of the protein in solution, with 
a time constant on the order of microseconds \cite{Malmendal1999,Evenas1999,
Evenas2001}, and was also suggested by theoretical and computational analysis 
of coarse-grained Hamiltonians \cite{Chen2007,Tripathi2009}.
Structural studies also supported the hypothesis that a small fraction of
calmodulin molecules can exist in a more compact conformation even in 
the presence of calcium \cite{Fallon2003}.

\begin{figure}[!h]
\begin{center}
\includegraphics[width=0.9\textwidth]{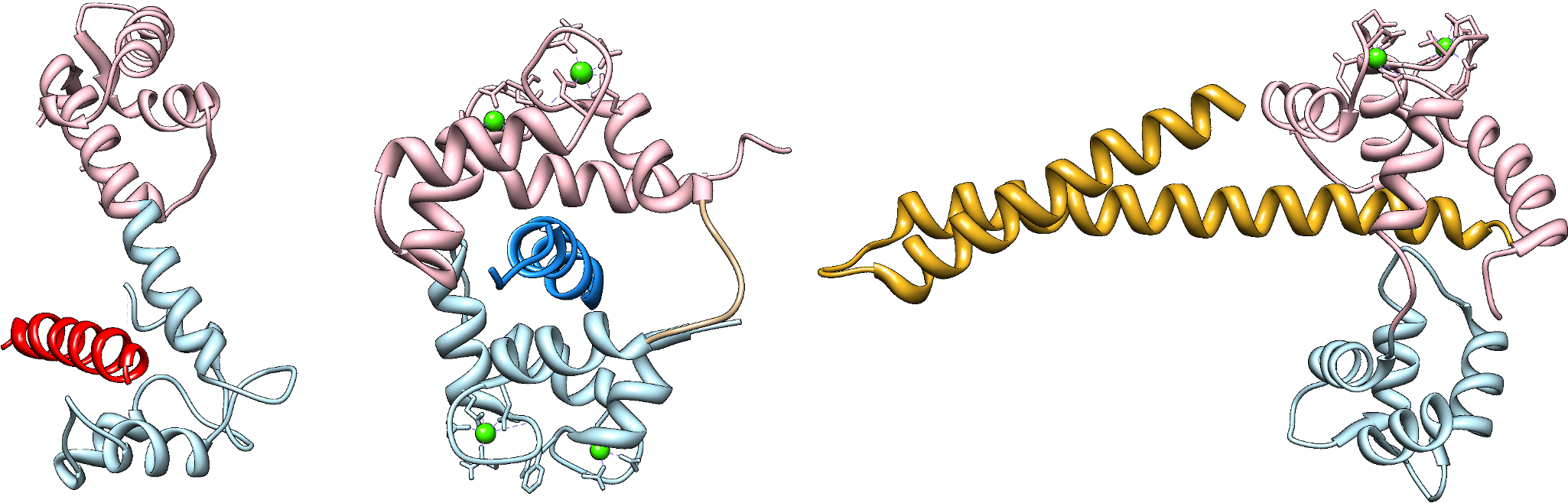}
\end{center}
\caption{
{\bf Examples of calmodulin's capability to bind diverse targets.} 
Calmodulin's N-lobe and C-lobe are in pink and light blue, respectively. Calcium ions are in green.
Left: calcium-free calmodulin binding to a peptide of neurogranin (red) with its C-lobe \cite{Kumar2013b}.
Center: calcium saturated calmodulin wrapping around a peptide of MLCK (blue) with both lobes \cite{Ikura1992}.
Right: calmodulin binding to a peptide of the SK channel (yellow) via a calcium-saturated N-lobe \cite{Schumacher2001}.
Several studies have shown that the linker region between the two domains is inherently flexible, and can assume different conformations depending on the orientation that the globular domains require for binding \cite{VanDerSpoel1996}.}
\label{fig:cam_binding_modes}
\end{figure}

Experimental studies with constrained mutants showed that the capability to 
switch conformation is necessary for cooperativity \cite{Tan1996,Meyer1996,Grabarek2005}, which is compatible with an MWC model.
Interestingly, conformational transitions are not common to all EF-hand
based proteins, despite their high structural homology in calcium-free
conditions.
For example, calbindin-28k is structurally nearly identical to a calmodulin
N-lobe, but its EF-hand remains closed after binding calcium \cite{Ababou2001}. 
It was proposed that the hydrophobic pockets of the open calmodulin 
lobes are stabilised by the unusual local abundance of the usually rare 
methionine residues (over 6\% for calmodulin, against the 1\% of the average protein) which have the highest flexibility, minimum steric hindrance and minimum solvation energy of all hydrophobic amino acids, and can therefore adapt to both segregated and and solvent-exposed conditions \cite{Nelson1998}.
The flexible aliphatic side chains of methionine can also easily establish 
contacts with very diverse substrates, contributing to calmodulin's promiscuity \cite{Ikura2006}.
Some examples of calmodulin's capability to bind very diverse targets is shown in Figure  \ref{fig:cam_binding_modes}.

\begin{figure}[!h]
\begin{center}
\includegraphics[width=0.9\textwidth]{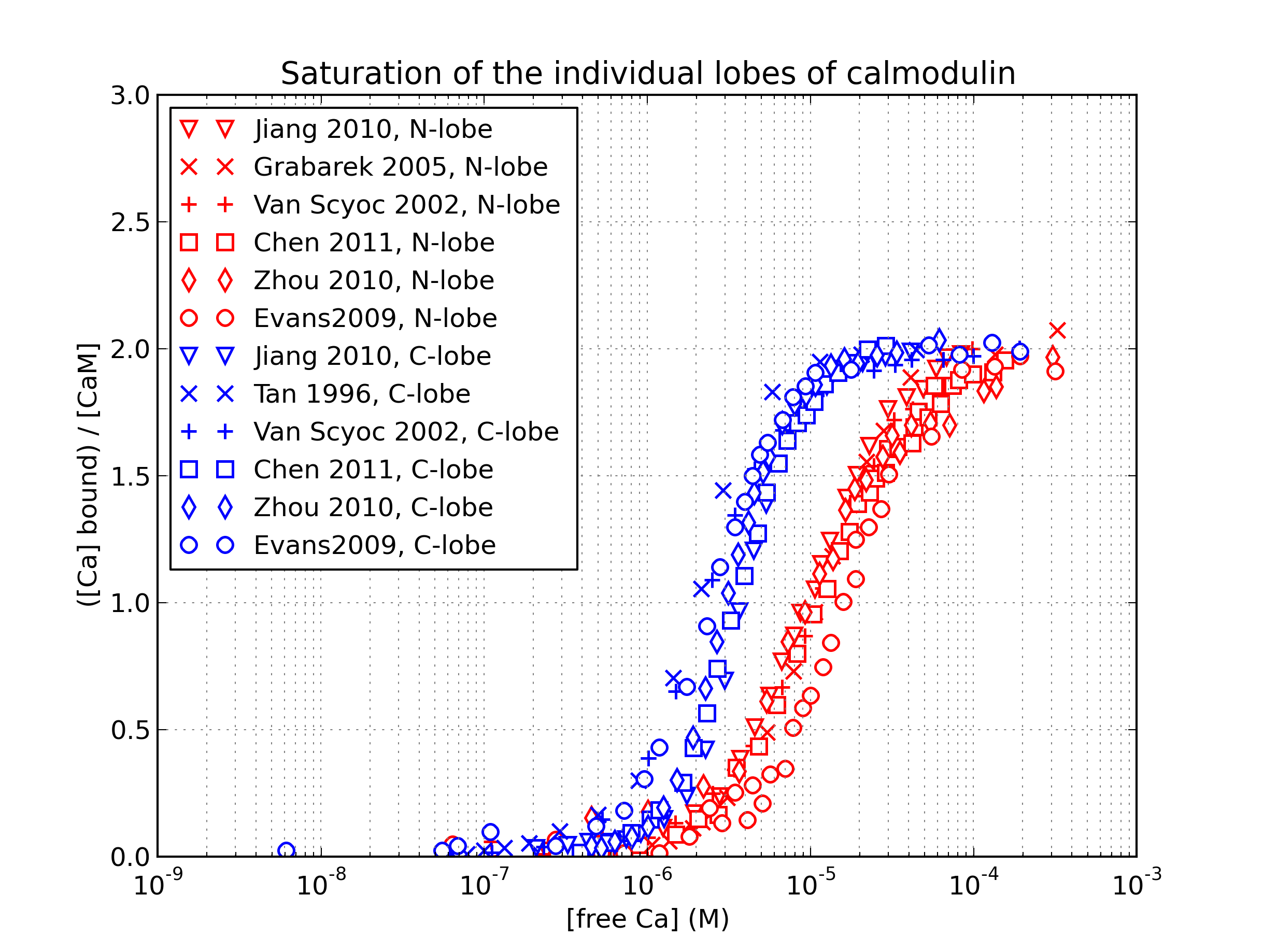}
\end{center}
\caption{
{\bf Different apparent calcium affinity of the N and C lobe.}  Experimental 
data was taken from references \cite{Jiang2010,Grabarek2005,VanScyoc2002,
Evans2009,Tan1996,Chen2011,Zhou2010}
}\label{N_vs_C}
\end{figure}

Despite their conceptual simplicity, the application of allosteric models can
be challenging. The intrinsic calcium affinities of different conformational
states are not easily measurable and need to be reverse-engineered, because
most experimental results are fitted using the phenomenological Hill or
Adair-Klotz models, which do not incorporate conformational transitions.
In addition, modelling intact calmodulin, within which each domain can switch
its conformation independently, results in a large number
chemical species that need to be explicitly enumerated (combinatorial
explosion), and a larger number of parameters that need to be fitted 
simultaneously.
A previous allosteric model of calmodulin was developed by Stefan et al.
\cite{Stefan2008}, to model the differential calmodulin-dependent activation 
of calcineurin and CaMKII in synaptic plasticity.
However, the model postulated that both lobes would undergo concerted
conformational transitions, and also had similar calcium-binding properties.
In fact, the two lobes possess a remarkable degree of autonomy \cite{Finn1995},
and the calcium saturation curve of calmodulin was amost exactly recovered by 
superposition of the saturation curves of tryptic fragments TR1C and TR2C, 
containing respectively the N-terminal or C-terminal lobe only
\cite{Minowa1984,Linse1991,Peersen1997}. 
The four alpha-helices of each globular domains form two EF-hands 
that work together as one cooperative unit with two calcium-binding sites 
\cite{Nelson2002}.
Despite of their high level of structural similarity, the two domains also
exhibit strikingly different affinities and binding kinetics for calcium ions. The
C-lobe has a 10-fold higher affinity, but much slower kinetics, than the N-lobe
\cite{VanScyoc2002,Faas2011}.
The clearly different saturation curves of the lobes, as observed in the intact
molecule, are shown in Figure \ref{N_vs_C}.

\begin{figure}[!h]
\begin{center}
\includegraphics[width=4in]{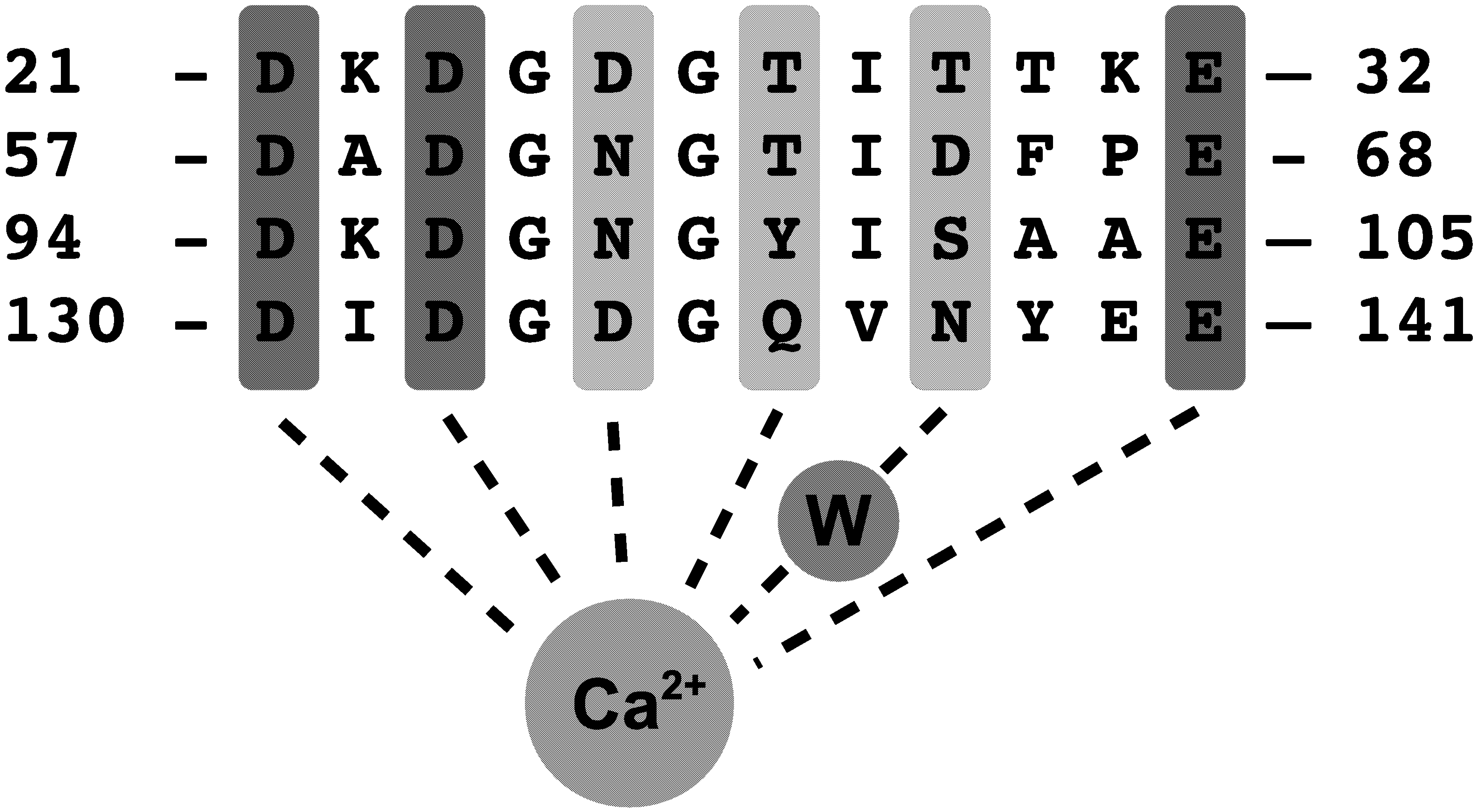}
\end{center}
\caption{
{\bf Sequence of the four calcium-binding pockets (EF-loops) of Calmodulin.}  
The calcium ion coordinates with the first,third,fifth,seventh and twelft
residue of the loop, and also with the ninth through a coordinating water
molecule (W). Highly conserved residues are in dark gray.}.
\label{fig:EF-loops}
\end{figure}

The differences in calcium-binding behavior are surprising in domains
that are structurally so similar, but several studies showed that the
EF-hand motif is tunable over a wide range of affinity and kinetics by
modification of few key residues \cite{Drake1996,Drake1997}. 
The incredibly high level of sequence conservation across species suggests 
that the different properties of the two lobes are some way crucial to 
calmodulin's function.
Moreover, even mutations that do not affect calcium-binding properties, 
but alter calmodulin's affinity for one or more targets, can be lethal
\cite{Wang2004}.
The sequence of calmodulin's four EF-loops (12-residue calcium-binding pockets
within each EF-hand) is given in Figure \ref{fig:EF-loops}.

The C-lobe was reported to play a pivotal role in mediating calmodulin's 
calcium-dependent interactions with its targets \cite{Kubota2007}, and
experiments with tryptic fragments of calmodulin consistently showed that
the affinity of the C-lobe for calmodulin-binding domains is usually much 
higher than the affinity of the N-lobe \cite{Bayley1996,O'Donnell2011}.
Moreover, targets that bind calcium-free calmodulin seem to interact almost
exclusively with the C-lobe \cite{Cui2003,Feldkamp2011,Chichili2013}.
In the light of these facts, we postulated that a major portion of the  
observed target-induced modulation effects could be explained by the
interactions of the targets with the C-lobe only, and we focussed our 
initial effort on modelling the observed properties of TR2C, (i.e. the 
isolated C-lobe).
Taking advantage of existing experimental data sets, we modelled the behavior of
the TR2C tryptic fragment  in the presence of peptides WFF and WF10 (peptides
that express full-length and truncated versions of the calmodulin-binding domain
of skMLCK) and Nav1.2IQp (a peptide based on the calmodulin-binding domain of
the calcium-activated NaV1.2 sodium channel). We developed and parameterised a
model of the tryptic fragment, in order to reliably reverse-engineer some intrinsic
properties of the C-lobe of calmodulin.  
We then parameterised an MWC model of the isolated N-lobe by postulating that 
the R-states of both lobes had very similar affinities (as suggested by 
\cite{Peersen1997}), thus reducing the number of free parameters to estimate 
(see Methods). 

This simplifying assumption was needed to circumvent the 
comparative scarcity of experimental data on the isolated N-lobe.
The two submodels were then combined into a model of intact calmodulin, under
the assumption that the protein behaved as the sum of its parts.

The MWC model explains cooperativity and target-induced affinity modulation 
as emergent properties, rather than \emph{a priori} assumptions, as opposed
for example to the Adair-Klotz model. 
The two formulations are however mutually consistent, and MWC and Adair-Klotz 
models are interconvertible, as shown by Stefan et al.
\cite{Stefan2009}, since for any MWC model the corresponding Adair-Klotz parameters can be computed.
A notable advantage of a MWC model is that the effects of multiple competing 
targets are straightforward to incorporate, simply by defining each target's 
affinity for the different conformational states of calmodulin.
The work described in this paper shows that a carefully parameterised 
MWC model can indeed give reliable predictions of how individual targets
modulate calmodulin affinity, and can also help investigate biologically relevant situations where numerous targets simultaneously modulate (and compete for) the same calmodulin pool.  

\section*{Results}

\begin{figure}[!h]
\begin{center}
\includegraphics[width=0.9\textwidth]{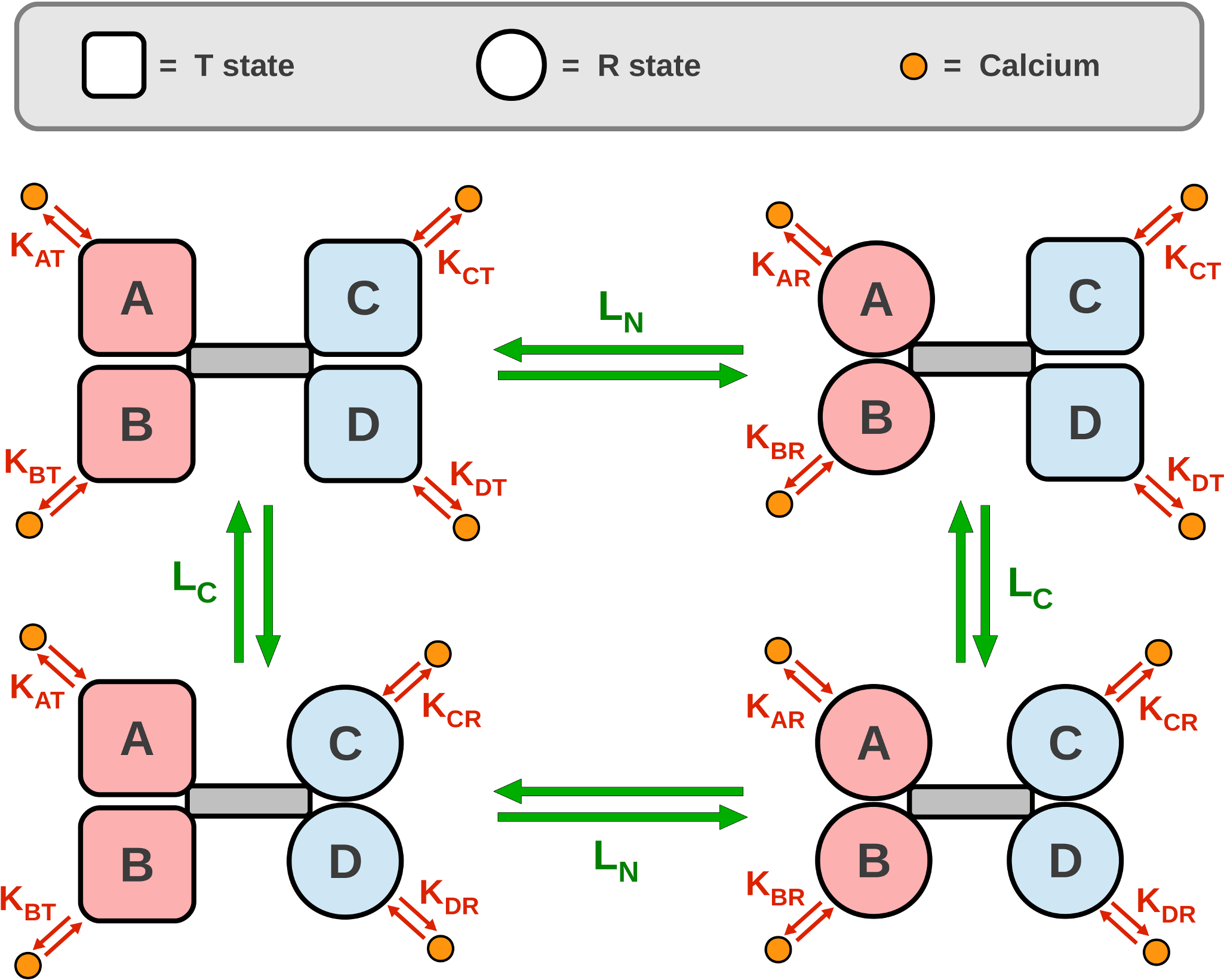}
\end{center}
\caption{ 
{\bf Allosteric model of calmodulin with hemiconcerted conformational transitions.}
Calmodulin has 4 calcium-binding sites (A, B, C, D) corresponding to one EF-hand each, and organised in two globular domains, N-lobe (pink) and C-lobe (light blue). Green arrows represent conformational transitions, red arrows represent calcium binding. Each binding site is modelled as a subunit that has access to two conformational states: open, high-affinity R state, and closed, low-affinity T state. Subunits on the same lobe are constrained to be in the same state, but the two lobes are allowed to be in different states. When no ligand is present, 4 possible configurations (TT, RT, TR, RR) coexist in thermodynamic equilibrium.
A lobe that undergoes a T-to-R transition increases its affinity for calcium. 
Under these assumptions, it follows from classical linkage theory that ligand binding shifts the conformational equilibrium towards the open form, by stabilising the higher-affinity R state. Cooperativity is predicted as an emerging property.}
\label{fig:hemiconcerted_diagram}
\end{figure}

A diagram of the allosteric model of calmodulin with hemiconcerted conformational transitions is shown in Figure~\ref{fig:hemiconcerted_diagram}.
Our model was built in three steps. First, we parameterised a model of TR2C, 
(isolated C-lobe), then parameterised a model of TR1C (isolated N-lobe), and
finally the two submodels were merged into a model of intact calmodulin.
(see Methods). The model of intact calmodulin (in SBML format) was deposited in BioModels Database \cite{BioModels2010} and assigned the identifier MODEL1405060000.

\subsection*{Preliminary analysis of possible parameterisations}
\begin{figure}[!h]
\begin{center}
\includegraphics[width=0.6\textwidth]{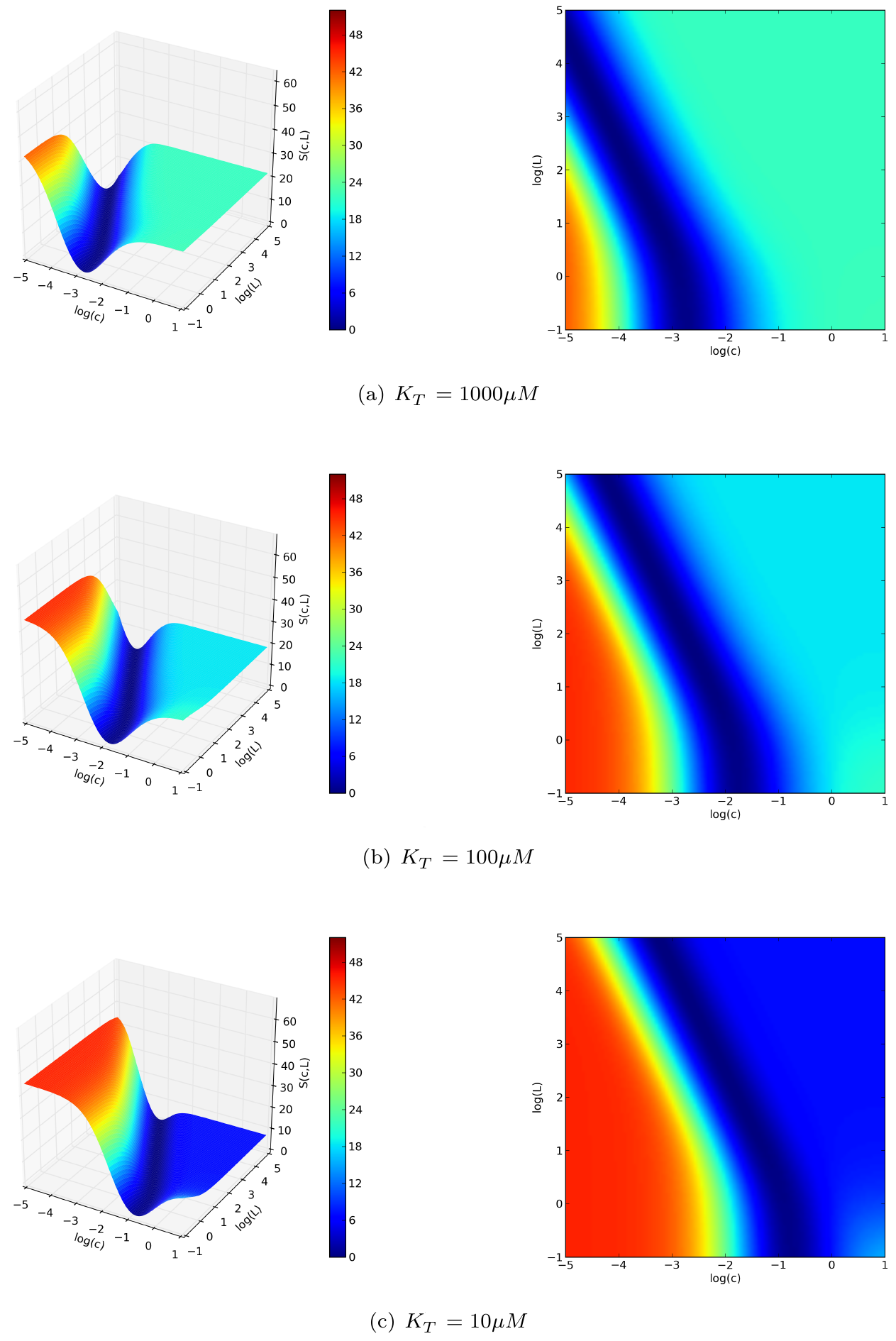}
\end{center}
\caption{ 
{\bf Exploratory study of the general beahviour of a two-site MWC model.}
The affinity of the T state was fixed at specific values within the estimated range and the sum of the squared
errors from the experimental points of the saturation curve, S, was plotted 
as function of the MWC parameters L and c. A lower score means a better fit. 
Over a very wide range of hypothetical, but physiologically plausible values 
of the T-state affinity ($1 mM - 10\mu M$), S(L,c) had the same qualitative
behavior. 
All the pairs of (L,c) values that lie at the bottom of a flat valley (dark blue area) give a very good fit of the saturation curve of TR2C in the absence of targets.
The knowledge of the target-free saturation function alone is therefore
insufficient for univocal parameter identification.}
\label{fig:fitting_landscapes}
\end{figure}

The first requirement we set for our model was the capability to
reproduce the saturation curve of  TR2C alone. In the absence of targets,
Equation~\ref{eq:rubin_changeux} depends on only 3 free parameters (L, $K_R$, $K_T$).
We observed that the intrinsic affinity of the T state must lie within
the wide, and biologically plausible, $1$~\textmu{}M - 1~mM range. Once the value
of $K_T$ is chosen, the model can be rewritten as a function of the
``classical'' MWC parameters, L and c, where $c = K_R / K_T$ \cite{Monod1965}.
For any choice of $K_T$, the score function of the fitting is a surface on 
the the ($L,c$) plane, with the same qualitative features, as shown in 
Figure~\ref{fig:fitting_landscapes}.

In particular, regardless of the value chosen for $K_T$, all the generated
fitting landscapes showed a flat valley, i.e. a region of parametric space where many
different choices of L and c could all fit the saturation curve very well.
Although it was technically possible to locate a point of global minimum in this
region, its position was not robust to noise, and moved erratically within the valley if
random errors were applied to the data sets (data not shown).
This portion of parameter space contained an ensemble of possible parameter
choices that lie between two limit cases:
in the first one, the change of affinity upon conformational transition is
relatively small ($c \simeq 0.01$), and the allosteric constant is also comparatively 
small($L \approx 100$); in the second one, the change of affinity is much greater 
($c \leq 0.001$) and so is the allosteric constant ($L \geq 1000$).
In the first case, (higher c, lower L) the molecule has a high propensity to 
spontaneously switch conformation to the high-affinity R state, but the R
state is stabilised less strongly by the binding of calcium.
In the second case (smaller c, greater L), the opposite holds true: the molecule
has a very low propensity to spontaneous conformational transitions, but the 
stabilising effect of calcium is much stronger.
It must be stressed that in both cases the model is capable of giving an
excellent fit of the experimental saturation curve in target-free conditions. 
The knowledge of the saturation curve in the absence of targets, alone, is therefore
insufficient to discriminate between the two possibilities and univocally identify the
parameters.
The problem could be solved using the additional information provided by the  
saturation curves observed in the presence of targets, which allowed us to 
constrain the parameter space to be sampled during the fitting procedure.
The obtained constraints implied that the first of the two limit scenarios 
mentioned above (with greater $c$ and lesser L) had to be be discarded, 
because the resulting model would be unable to account for the extent of 
the target-induced affinity shifts observed in the experiments, as explained
in the Methods section.
\subsection*{Truncation does not strongly affect the C-lobe's calcium affinity}

\begin{figure}[!h]
\begin{center}
\includegraphics[width=0.9\textwidth]{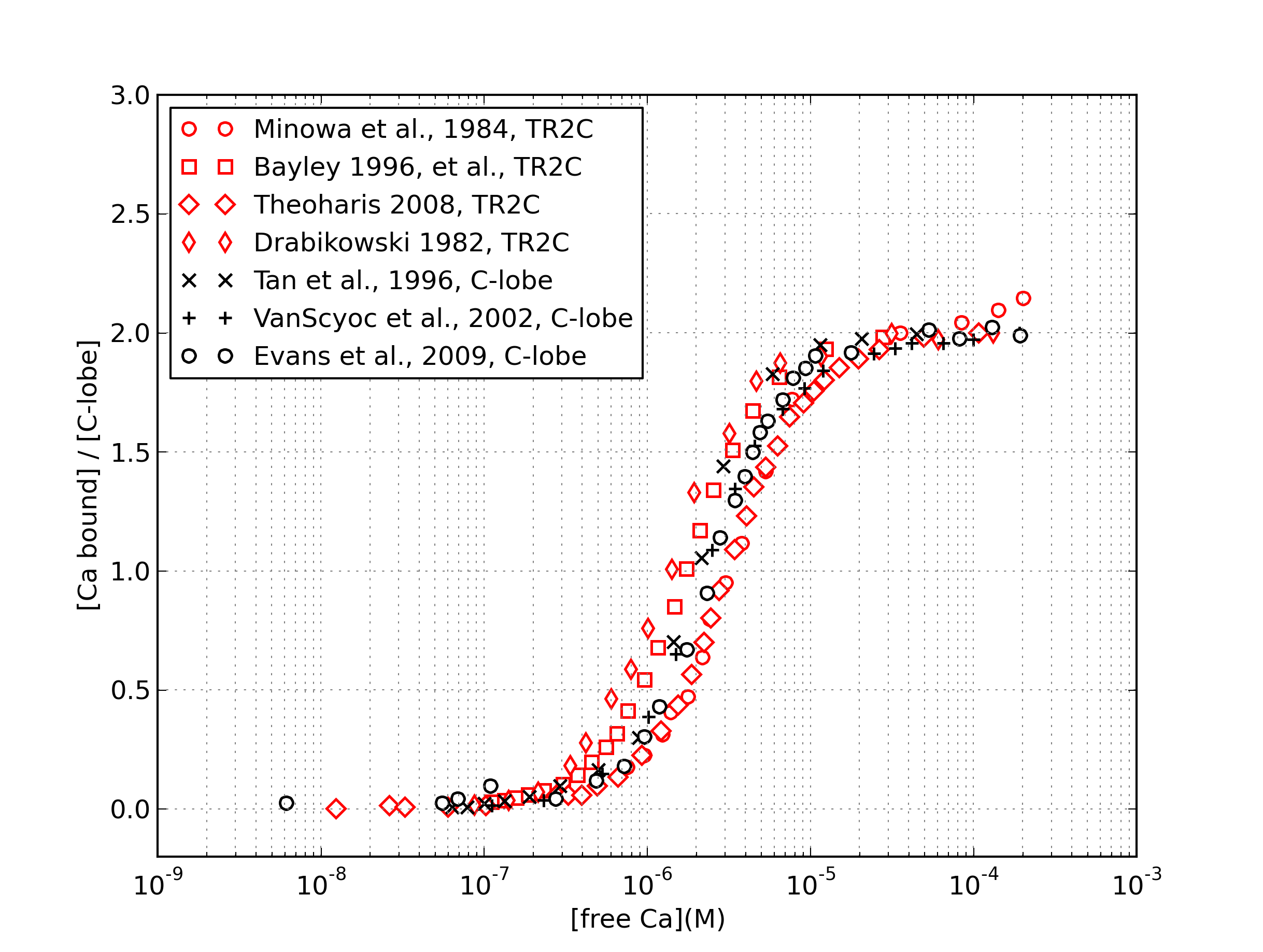}
\end{center}
\caption{
{\bf Comparison between the experimental saturation curves of TR2C and the C-lobe}. 
Saturation of TR2C (red) and C-lobe of intact calmodulin (black), as a function of
the free calcium concentration, monitored by intrinsic tyrosine fluorescence 
(data taken from references \cite{Drabikowski1982,Minowa1984,Tan1996,Bayley1996,Theoharis2008,Evans2009, VanScyoc2002}). 
The available experimental data show some scattering, with an uncertainty of
about a factor two for the concentration of free calcium necessary to produce
half-saturation.}\label{fig:tr2c_vs_ccam}
\end{figure}

We tested whether it was legitimate to reverse engineer the intrinsic properties
of the C-lobe from those of TR2C.
The level of saturation of the C-lobe can be monitored in both intact calmodulin
and TR2C fragments by measuring the intrinsic fluorescence of their tyrosine
residues \cite{VanScyoc2002,Johnson1996,Theoharis2008,Feldkamp2011,Evans2009}, 
It was shown that truncation does not have a strong effect on the secondary 
structure and tyrosine fluorescence intensity of the C-lobe
\cite{Drabikowski1982}.
In order to verify that the calcium-binding properties of the intact
calmodulin's C-lobe were maintained in the TR2C tryptic fragment, we compared
the two saturation curves as presented in Figure~\ref{fig:tr2c_vs_ccam}. 
The data shows some scattering, and the uncertainty on the calcium concentration
necessary to produce half-saturation is roughly of a factor two. This is
probably due to slightly different experimental conditions (see the Discussion
section). 
However all curves were similar and shared the same qualitative behavior,
meaning that truncation does not dramatically alter the lobe's properties.
\subsection*{Simultaneous fitting of saturation curves with and without
peptides}
We estimated the model parameters by fitting experimental calcium saturation 
curves of TR2C alone, TR2C+WFF peptide (1:1.4 molar ratio), TR2C+Nav1.2IQp 
peptide (1:1.4 molar ratio). A fourth and fifth data set, TR2C+WF10 (1:1.4 
molar ratio) and TR2C+Nav1.2IQp peptide (1:2.8 molar ratio) were not used for 
parameter fitting and were kept for validation purposes. Data sets were taken 
from references \cite{Bayley1996, Theoharis2008}.

In the absence of targets, the COPASI model contains 3 independent parameters, 
(L,$K_R$,$K_T$). In the presence of target $t$, the saturation curve depends 
also on the target affinities $K_r^t$, $K_T^t$, which were both known quantities 
for WFF.
The affinities of the peptide NaV1.2IQp for both the T and R sates were in 
the nanomolar range, and only estimated upper limits for their value were
available in the literature (Table~\ref{tab:peptides_summary}). 
We assumed that NaV1.2IQp affinity for the R state was equal to the estimated
upper limit, while its affinity for the T state was set as an 
additional parameter to be fit, which we called $K^{NaV}_T$. 

In total, the 4 independent parameters $K_T$, $K_R$, $K^{NaV}_T$ and
$L$ were fit simultaneously to the three experimental data sets.

\begin{figure}[!h]
\begin{center} 
\includegraphics[width=0.9\textwidth]{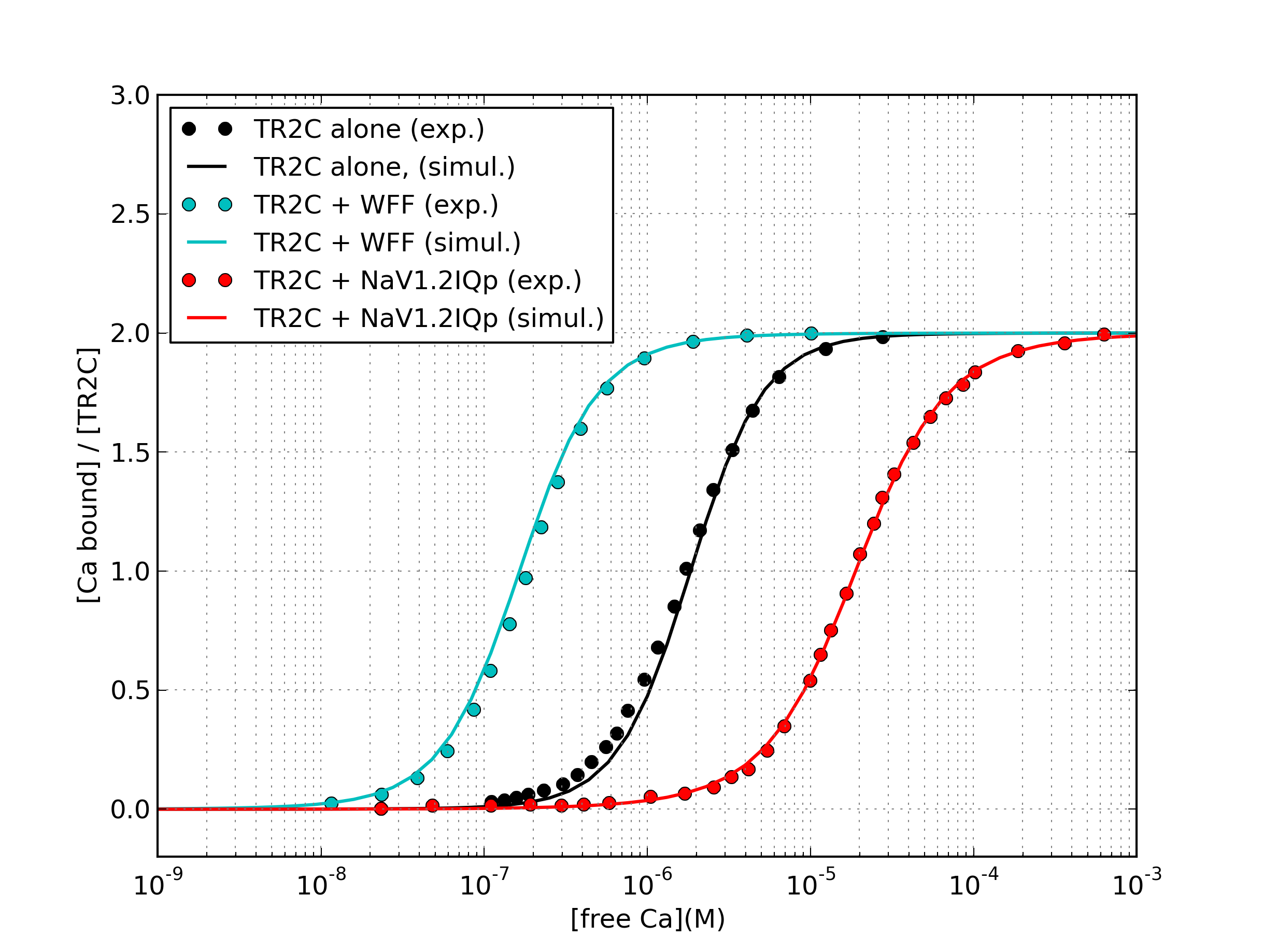}
\end{center}
\caption{
{\bf Comparison between fitted and experimental saturation curves.}
Experimental data was taken from references \cite{Bayley1996} and 
\cite{Theoharis2008}.}\label{fig:saturation}
\end{figure}

The estimated parameters are summarised in Table~\ref{tab:fitted_parameters},
with the resulting saturation curves shown in Figure~\ref{fig:saturation}.
The simulations of the model were overall in good agreement with experiments.

\begin{table}[!h]
\begin{center}
\begin{tabular}{l l l l l}
\hline
&&&&\\
Peptide 	& Description 	&  $K^t_T$  & $K^t_R$ 	& Source  \\
\hline
&&&&\\
WFF		& Full-lenght CaM-binding domain of skMLCK	& $1.6 \mu M$ 
& $76 nM$ 	& \cite{Bayley1996} 	\\
&&&&\\
WF10		& Truncated CaM-binding domain of skMLCK 	& $1.1 \mu M$ 
& $712 nM$ 	& \cite{Bayley1996} 	\\
&&&&\\
NaV1.2IQp	& IQ domain of the NaV1.2 sodium channel	& $\leq 25 nM$ 
& $\leq 75 nM$ & \cite{Theoharis2008}	\\
\hline
\end{tabular}
\end{center}
\caption{Summary of the peptides used in the experiments described in
references \cite{Bayley1996,Theoharis2008, Evans2009}, and their affinities for the
truncated C-lobe of CaM in the T and R state. The superscript $t$ indicates calmodulin's 
affinity for a target. These peptides were used for the fitting and
validation of the computational model.}\label{tab:peptides_summary}
\end{table}

\begin{table}[h]
\begin{center}
\begin{tabular}{l c l l }
\hline
& & &  \\
Parameter & (Units) & Value & $\pm$ Std\\
\hline
& & &  \\
$L$ 	& /	&  $8.616\cdot 10^{3}$  & $\pm$ $4.961\cdot 10^{3}$  \\
& & &  \\
$K_T$ 	& M  	&  $6.241\cdot 10^{-5}$ & $\pm$ $5.839\cdot 10^{-6}$ \\
& & &  \\
$K_R$   & M	&  $1.979\cdot 10^{-8}$	 & $\pm$ $5.631\cdot 10^{-9}$ \\
& & &  \\
$K^{NaV}_T$ 	& M	&  $6.095\cdot 10^{-10}$& $\pm$ $7.979\cdot 10^{-11}$\\
\hline
\end{tabular}
\end{center}
\caption{Summary of the estimated parameters for the model of TR2C. The reported standard
deviations are automatically computed by COPASI, by inversion of the Fisher
information matrix ofthe score function.}
\label{tab:fitted_parameters}
\end{table}

\subsection*{Validation}
\begin{figure}[!h]
\begin{center}
\includegraphics[width=0.9\textwidth]{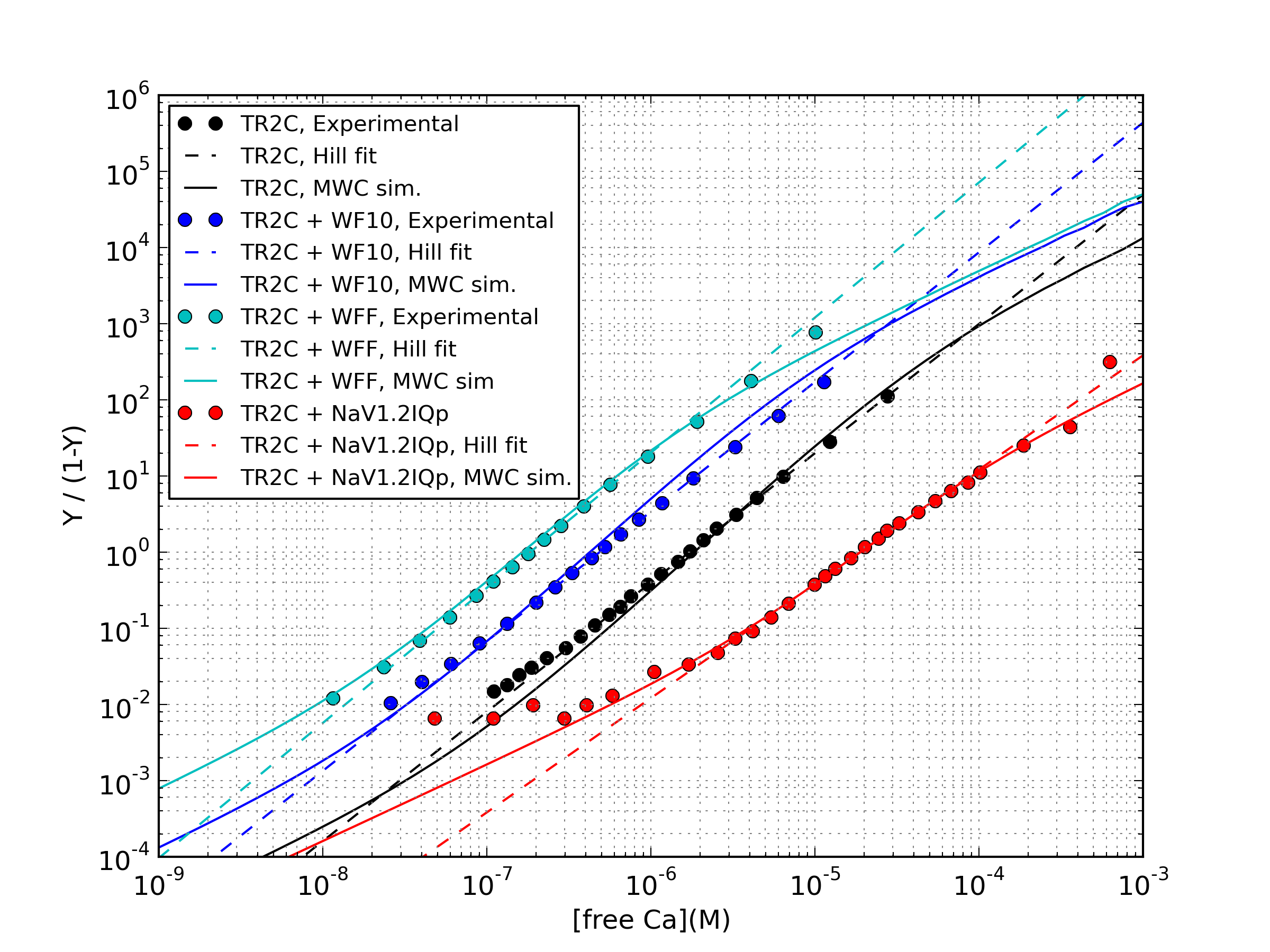}
\end{center}
\caption{
{\bf Hill plot summarising the predicted saturation curves of TR2C} 
in the presence of WFF, WF10 and NaV1.2IQp peptides, and comparison 
with the artificial datasets reverse-engineered from reference 
\cite{Bayley1996}, and actual experimental data from reference
\cite{Theoharis2008}. In the range of physiologically plausible calcium concentrations (from nanomolar to tens of micromolar) the MWC model (continuous lines) is close to the Hill model (dash lines).}\label{fig:summary_hillplot}
\end{figure}

Once the parameter values were determined, we tested the capability of the model to predict the behavior of TR2C under conditions different from those used for the fitting. The saturation curve of TR2C, as measured by Bayley and coworkers in the presence of WF10 peptide, (a truncated form of WFF with a 10-fold lower affinity for the R-state) was in good agreement with experiments. Moreover, the saturation curves by Evans and coworkers \cite{Theoharis2008} were measured with two different concentrations of peptide, and it was found that doubling the concentration of Nav1.2IQp (from 1:1.4 to 1:2.8 TR2C:peptide ratio) did not produce a further shift in the saturation curve, a behavior that our model was able to reproduce (see Supporting Information).%
A summary of all fitted and predicted saturation curves, plotted against the 
corresponding experimental data points, is given in
Figure~\ref{fig:summary_hillplot}.

\subsection*{The interplay of competing targets determines the affinity curve shift}
\begin{figure}[!h]
\begin{center}
\includegraphics[width=0.9\textwidth]{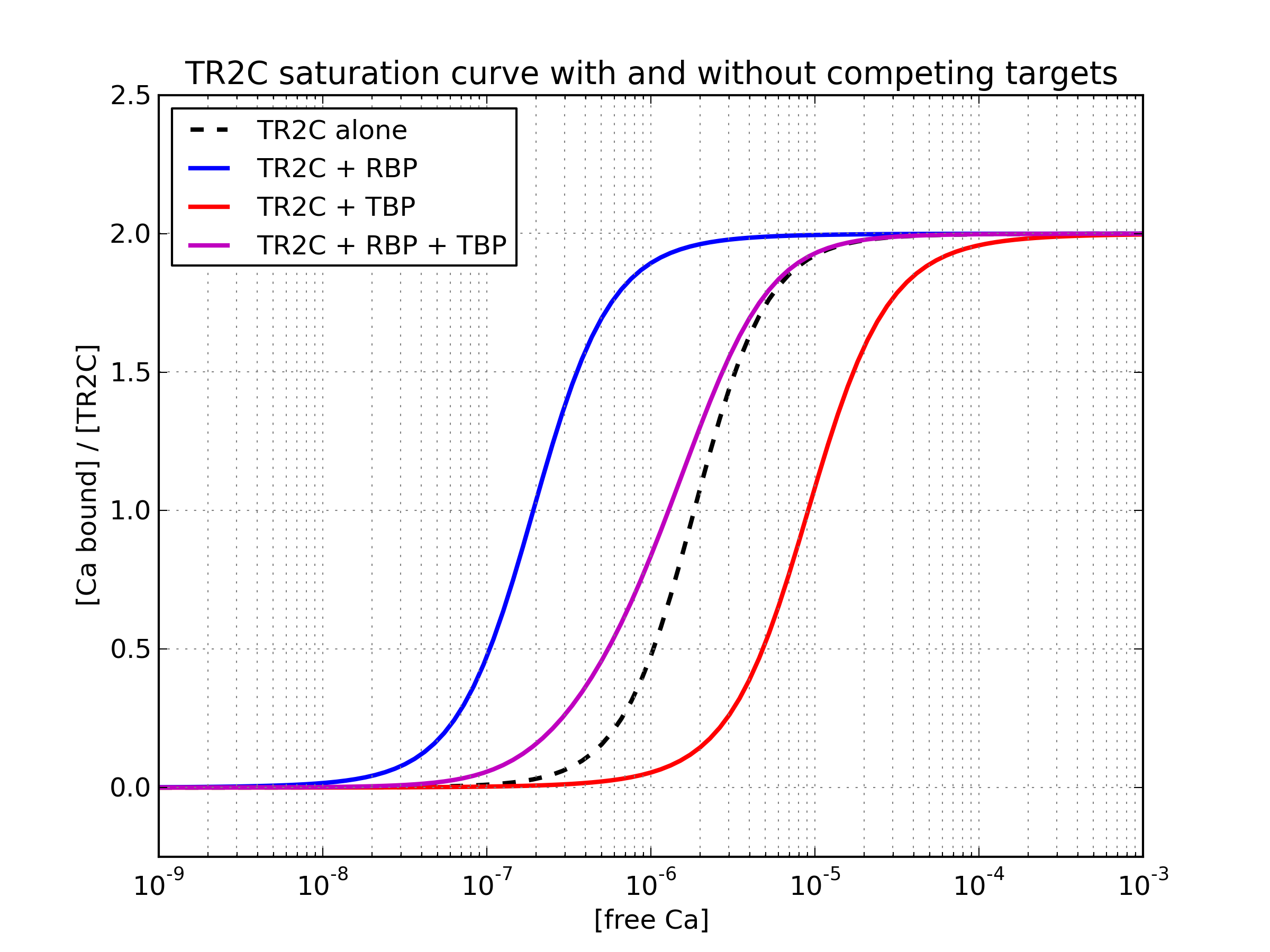}
\end{center}
\caption{ 
{\bf Cumulative effect of competing targets on TR2C saturation curve}, in a 
simulated reaction chamber with $40 \mu M$ TR2C, $40 \mu M$ of a target with
preference for the T-state (T-state binding target, TBT), and $100 \mu M$ of a
target with preference for the R state (R-state binding target, RBT). Each target alone can produce a marked shift in the saturation curve, but when both target are present in equal concentrations, the shift is much smaller. As targets, we chose peptides of two abundant neuronal proteins, neurogranin (Ng) and CaMKII, whose binding constants for TR2C were available in the literature. The chosen molar ratio (1:1:2.5) reflects estimated relative abundance of CaM, Ng and CaMKII in neuronal compartments.
Their affinities for the T and R state of TR2C were respectively: $43nM$ and $1.05\mu M$ for neurogranin, and $88 \mu M$ and $0.95 \mu M$ for CaMKII (\cite{Kumar2013b, Evans2009}).}\label{fig:competing_targets_saturation}
\end{figure}

The high calcium affinity of the C-lobe, further enhanced by interactions within some targets (CaMKII, PP2B, skMLCK), has led to speculation that calmodulin could activate such targets even at resting calcium concentrations, i.e. in the absence of calcium signals in neuronal or muscular cells \cite{Peersen1997}.
We used the model of TR2C to perform a preliminary investigation of the effect of competing targets on calmodulin. At this stage of analysis we can only investigate the part of such interactions that are mediated by the
C-lobe. However, as previously mentioned, the C-lobe was shown to be mainly 
reponsible for mediating calmodulin-target interactions, and targets that bind the T-state with high affinity, in particular those that appear to have very little interaction with the N-lobe \cite{Kubota2007}. 
We simulated the steady-state response of a system containing  calmodulin and two competing allosteric targets using data taken from the literature, 
one binding prefentially to the the T state (T-state binding target, TBT) and the other binding preferentially to the R state (R-state binding target, RBT).
When both targets are present, the total saturation curve depends on their
relative concentrations, as shown in Figure~\ref{fig:competing_targets_saturation}. Shifting the concentration of targets is a potential way to tune the saturation curve of the calmodulin pool.

\subsection*{Parameter estimation for the N-lobe}
The model of the isolated N-lobe, in the absence of targets, was based on the model of C-lobe. The resulting parameters are summarised in Table \ref{tab:nlobe_parameters}. For the detailed procedure, see the Methods section.
\begin{table}[h]
\begin{center}
\begin{tabular}{l c l l }
\hline
& & &  \\
Parameter & (Units) & Value & $\pm$ Std\\
\hline
& & &  \\
$L_N$ 	& /	&  $3.226\cdot 10^{5}$  & $\pm$ $2.486\cdot 10^{5}$  \\
& & &  \\
$K_T^N$ & M &  $9.192\cdot 10^{-5}$ & $\pm$ $5.278\cdot 10^{-5}$ \\
& & &  \\
$K_R^N$ & M	&  $1.979\cdot 10^{-8}$	 & $\pm$ $5.631\cdot 10^{-9}$ \\
\hline
\end{tabular}
\end{center}
\caption{Summary of the estimated parameters for the isolated N-lobe. 
The R-state affinity was assumed to be equal to that of the C-lobe.
The reported standard deviations were computed from the covariance matrix 
of the fit}.
\label{tab:nlobe_parameters}
\end{table}
\subsection*{Model of intact calmodulin}
\begin{figure}[!h]
\begin{center}
\includegraphics[width=0.9\textwidth]{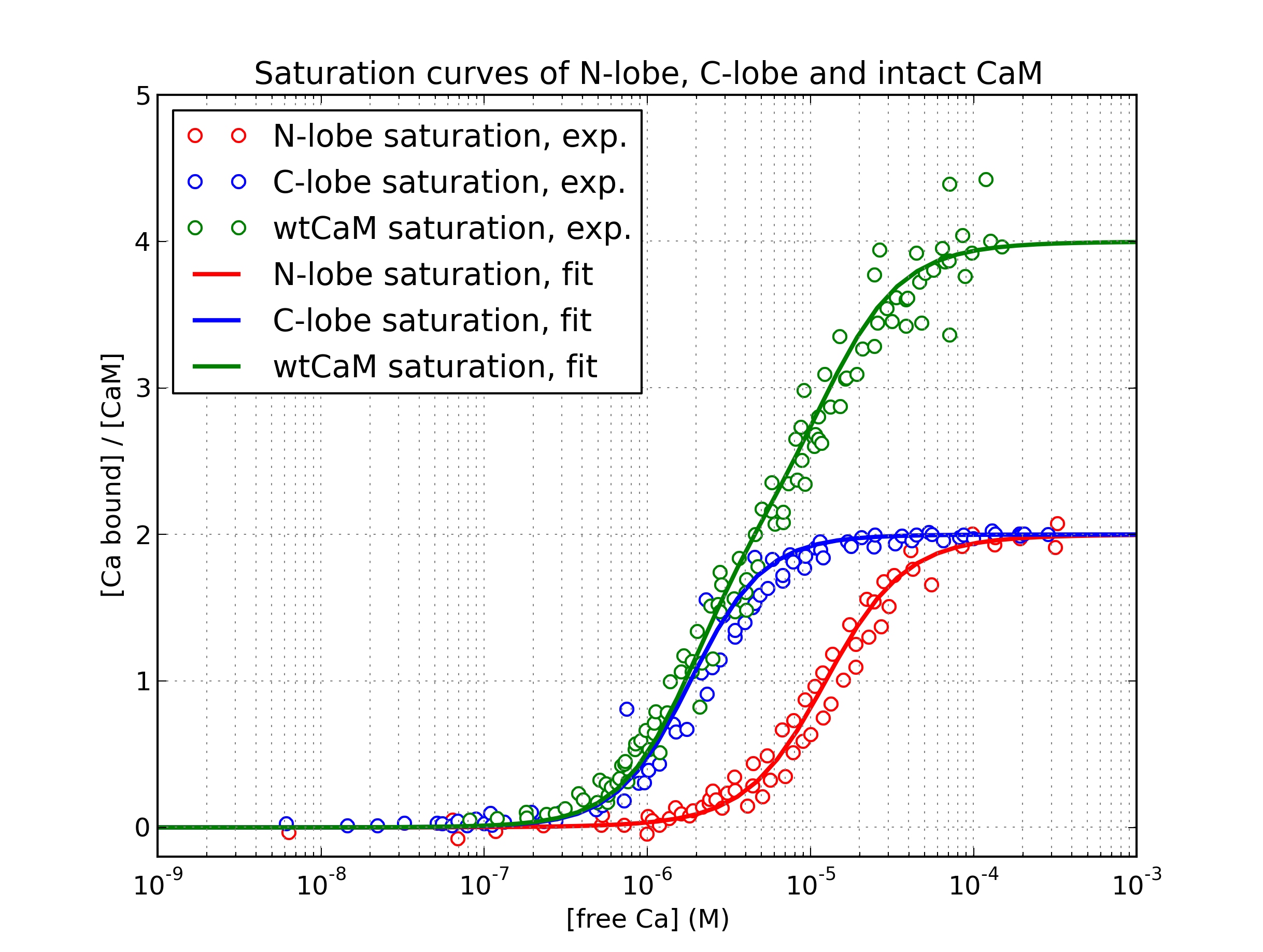}
\end{center}
\caption{ 
{\bf Saturation curve of individual lobes and intact calmodulin}, 
as predicted by our model, and comparison with experimental data.
The datapoints for intact calmodulin were taken from references
\cite{Porumb1994,Bayley1996,Peersen1997}.}
\label{fig:wtcam_n_c_saturation}
\end{figure}

The model of intact calmodulin was built by combining the two models of the N and C and lobe. The saturation curve agreed very well with the available experimental data as shown in Figure \ref{fig:wtcam_n_c_saturation}.
A model including calmodulin and binding targets was then automatically generated in SBML format, as described in the Methods section, and simulated in COPASI.

\subsection*{Modulation of intact calmodulin by targets}
We validated the model of intact calmodulin by fitting to available experimental data.
The effect of a target on the saturation curve of a MWC molecule depends on how the different conformations are stabilised, which in turn depends on the affinity of the target for each conformation. 
The allosteric models of isolated lobes assume that each lobe can only exist in two states, T and R, which implies that both EF-hands on each lobe always undergo concerted conformational transitions. One the other hand, the model of intact calmodulin contains two lobes but does not constrain them to be the same state. As a result, one needs to take into account "asymmetric" conformations, where the two lobes are in different states.
The model of intact calmodulin thus contains 4 possible states, called RR, RT, TR, TT. The first and second capital letter refer to the state of the N and C lobe, respectively. For more details, see the Methods section.

Bayley and coworkers measured calmodulin saturation in the presence of two peptides: WFF, which represent the full-length CaM-binding domain of skMLCK, and WF10, a truncated version of the same domain, corresponding to the portion that interacts with the C-lobe \cite{Bayley1996}. The saturation curve with WFF was markedly shifted to the left, while that in the presence of WF10 was biphasic.

\begin{figure}[h]
\begin{center}
\includegraphics[width=0.9\textwidth]{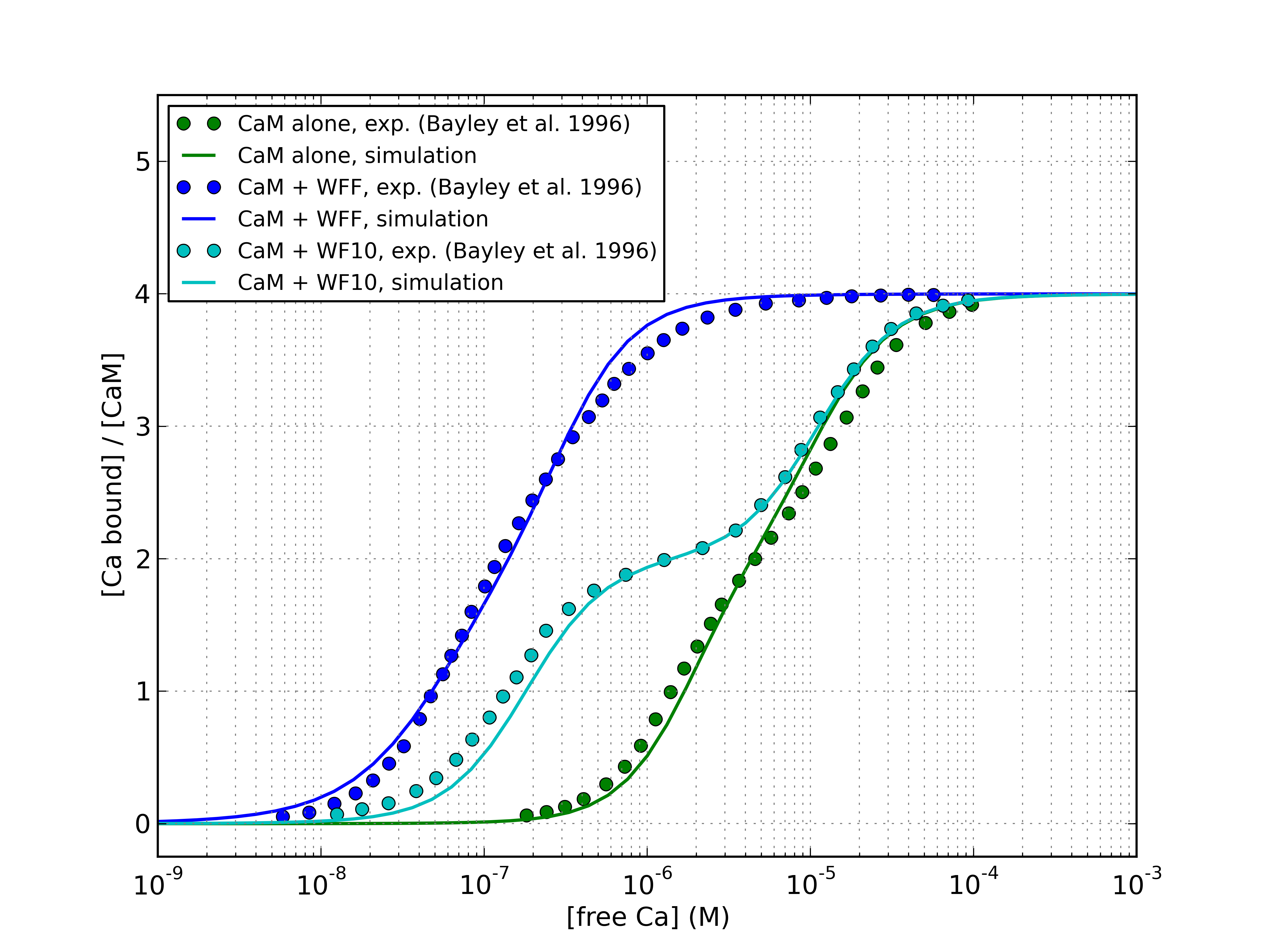}
\end{center}
\caption{ 
{\bf Effect of skMLCK peptides on the saturation curve of calmodulin}, 
as predicted by our model, and comparision with experimental data by \cite{Bayley1996}.}
\label{fig:cam_wff_wf10}
\end{figure}

Crucially, they were careful to measure the affinity of calmodulin and TR2C for these peptides, both in calcium and calcium-saturated conditions, which gave us an excellent initial estimate for the affinity of the peptides for the RR and TT state of intact calmodulin, and for the R and T states of TR2C. 
The affinity of each target for the isolated C-lobe in the R state was used as an estimate of the affinity of the target for the intact molecule in the TR state (i.e. we assumed that the target interacted with the C-lobe was much more strongly than with the N-lobe, in agreement with available experimental data).
The estimated affinities for the four states are summarised in Table \ref{tab:cam_target_affinities}.
The agreement between simulation and experiment was very satisfactory (Figure \ref{fig:cam_wff_wf10}).
\begin{table}[!h]
\begin{center}
\begin{tabular}{l l l l l}
\hline
&&&&\\
Peptide 	& $K^t_{RR}$  & $K^t_{RT}$    & $K^t_{TR}$ & $K^t_{TT}$  \\
\hline
&&&&\\
WFF		    & $<0.1 nM$ & $>600\mu M$ & $735nM$  & $>600\mu M$ 	\\
&&&&\\
WF10		& $735nM $  & $ 200\mu M$ & $735nM$  &  $ 200\mu M$ \\
\hline
\end{tabular}
\end{center}
\caption{Estimated affinities of WFF and WF10 peptides for the different conformations of intact calmodulin, based on assumptions and actual affinity measurements from reference \cite{Bayley1996}. The affinity of WFF for the RR state was only given as upper limit ($<0.1nM$), and was assumed to be $0.05nM$. The affinity of both peptides for the TR state was assumed to be equal to the peptide affinity for the isolated C-lobe in the R state. In the case of WF10, which is supposed to only interact with the C-lobe, the peptide affinity for the RR and TR states were set to be equal. Each peptide's affinity for the RT conformation (which is expected to be underpopulated in steady-state conditions and therefore play a marginal role in this analysis) was set equal to the affinity for the TT state.}
\label{tab:cam_target_affinities}
\end{table}
\section*{Discussion}
\subsection*{Calmodulin modulation by individual targets}
We have shown that allosteric regulation can explain the modulation of the calcium-binding properties of the TR2C fragment (and hence, of the C-lobe of calmodulin) by several calmodulin-binding peptides. 
The immediate consequence is that for a given concentration of target, the parameter that determines the extent of the saturation curve shift, at steady-state, is the ratio of the affinities of the target for the R and T states of calmodulin.
Plotting the experimental points against the fitted curves in the Hill plane also offer an explanation as to why the MWC and Hill models can both provide a very good fit, despite predicting two quite different qualitative behaviors: in the Hill plane, the Hill curve is a straight line, whilst the MWC model predicts a gradual shift from a lower to a higher asymptote.
As shown in Figure~\ref{fig:summary_hillplot}, in the range of physiologically plausible calcium concentrations (from nanomolar to tens of micromolar) the MWC model is close to the Hill model. The Hill coefficient predicted by the MWC model also decreases in the presence of targets, in qualitative agreement with experiments (Table~\ref{tab:hill_coefficients}).
The simulations of intact calmodulin in the presence of WF10 shows that biphasic saturation curves can be produced by targets that only stabilise the high-affinity form of one lobe. WF10 is an artificial peptide, but targets that induce differential modulation of calmodulin domains can be found in nature, as in the case of the anthrax edema factor \cite{Ulmer2003}.

An allosteric model of calmodulin was previously developed in our group that could satisfactorily reproduce the dose-response curve of wildtype calmodulin and also explain differerential activation of PP2B and CaMKII during synaptic tetanic stimulation \cite{Stefan2008}.
The previous model was however based on different premises and some additional simplyfying hypotheses, such as concerted conformational transitions of both lobes, and exclusive binding of target by one conformation.
A direct comparison would therefore not be meaningful.

\begin{table}[!h]
\begin{center}
\begin{tabular}{l l l}
\hline
&   \\
Case & $n_H$ (MWC) &  $n_H$ (Hill) \\
\hline
&   \\
TR2C 			& $1.91$ &   $1.71$\\
&   \\
TR2C + WF10		& $1.89$ &   $1.70$\\
&   \\
TR2C + WFF  		& $1.77$ &   $1.77$\\
&   \\
TR2C + NaV1.2IQp	& $1.53$ &   $1.50$ \\
&   \\
\hline
\end{tabular}
\end{center}
\caption{Summary of Hill coefficients for TR2C in the absence and presence of
peptides. Values were calculated as the maximum slope of the Hill plot of the MWC model
(left column) or from fitting the experimental saturation curve with a Hill equation (right
column).}\label{tab:hill_coefficients}
\end{table}

\subsection*{Calmodulin modulation by competing targets}
Calmodulin \textit{in vivo} is always in the presence of a large number of 
targets, which simultaneously modulate its activity.
We used our model to investigate the effect of mixtures of competing targets 
on calmodulin's C-lobe. 
We exploited the seemingly predominant role of the C-lobe in mediating 
calmodulin-target interactions to test the behavior of our model in 
the simultaneous presence of two different targets that had higher affinity 
for the T state and the R state, respectively.
As an example scenario we chose peptides of abundant neuronal proteins, whose 
binding affinities for TR2C were available in the literature.
We chose a 1:1:2.5 molar ratio for the three proteins, (a scenario where the 
concentration of calmodulin-binding proteins is much higher than the concentration
of calmodulin \cite{Persechini2002}), to account for the higher concentration 
of targets that favour the R-state.
In the chosen conditions, the effect of the competing targets was almost 
cancelled out and the resulting calcium saturation curve was close to that 
in the absence of targets, as shown in Figure \ref{fig:competing_targets_saturation}. 
Regulating the relative abundance of targets can therefore bidirectionally tune 
the amount of calcium bound to calmodulin's C-lobe.

\subsection*{Two-state approximation and parameter estimation}
\begin{figure}[!ht]
\begin{center}
\includegraphics[width=0.9\textwidth]{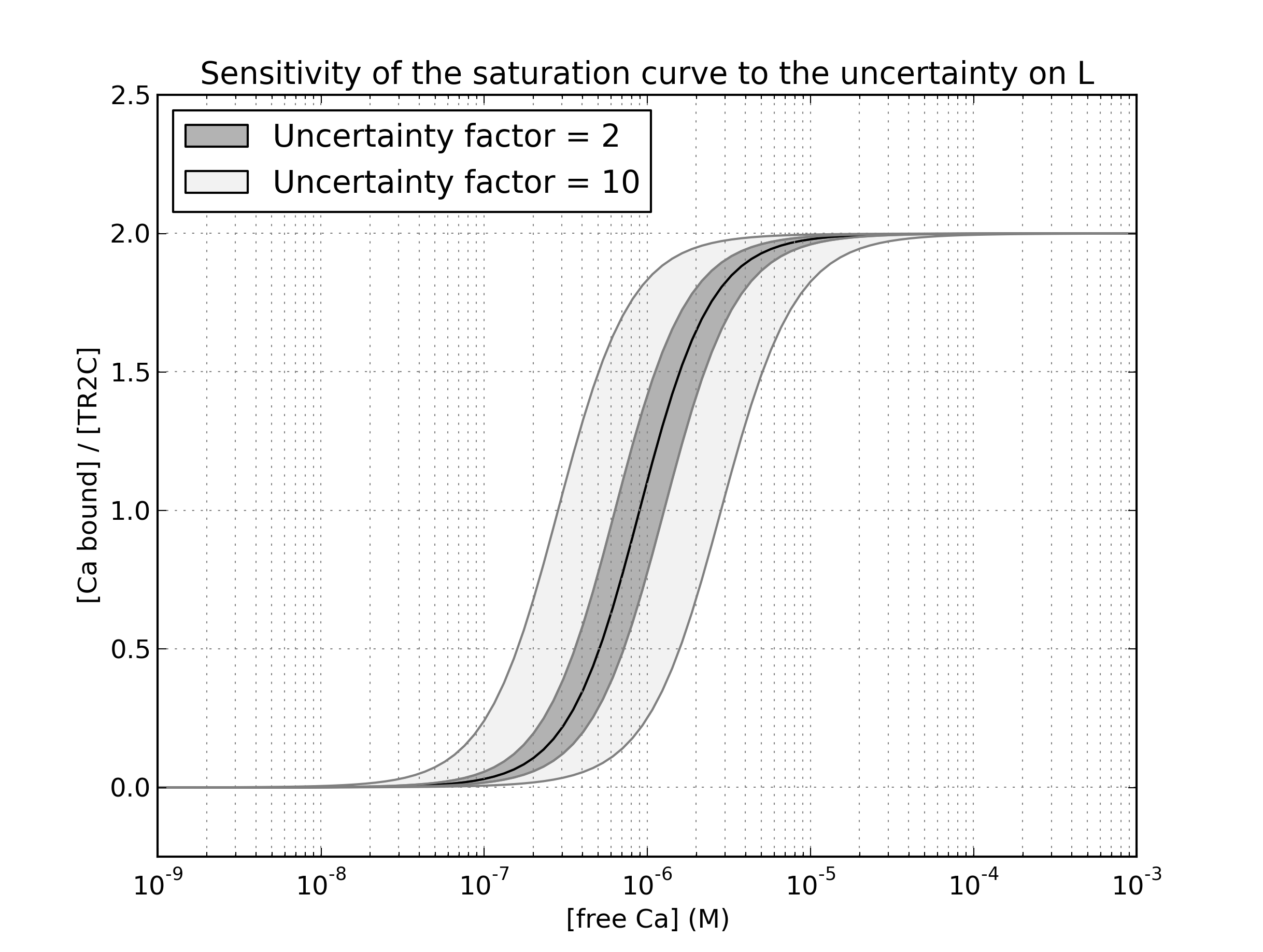}
\end{center}
\caption{
{\bf Sensitivity of the saturation curve to the value of the allosteric constant.} 
The black line is the fitted function, the grey area shows the region
that the curve would span if the allosteric constant was lowered (left shift) or
increased (right shift) by a factor two (dark gray) or 10 (light gray).}\label{fig:l_sensitivity}
\end{figure}

A two-state model is of course a simplification of something as complex as a 
protein's conformational dynamics. It is more plausible that calmodulin is capable of sampling a wider ensemble of conformations, and its high conformational plasticity allows it to regulate downstream protein targets that are structurally very diverse \cite{Feldkamp2011,Ikura2006,Kumar2013a}. 
Moreover, the PEP-19 peptide, expressed in the cerebellum, was shown to regulate mostly calmodulin's calcium binding kinetics, with little effect on affinity, and could do so even when calmodulin was already associated to CaMKII \cite{Putkey2003}.
These observations imply that the tuning of calmodulin's affinity and kinetics is highly adaptable, but capturing every possible feature of this mechanism was outside the purpose of the present work. Moreover, for the conditions focussed on here, the agreement of a two-state model with experimental results was already satisfactory. 
On a technical level, parameter estimation is made challenging by the uncertainties and the scattering shown by the available experimental data.
Ideally, allosteric models would have to be parametrised using data sets 
produced in highly standardised conditions, designed to be as close to the
\textit{in vivo} conditions as possible. However, different groups performed similar experiments and measured apparent calcium affinities that sometimes differ by up to a factor of two or more. This could be due to a number of reasons, including variable levels of purity of reagents or proteins, or slightly different buffer conditions. For example, increasing KCl concentration from 92 to 152mM was reported to decrease
the apparent calcium affinity by a comparable factor \cite{Pedigo1995}. As a
consequence, pinpointing the exact value of some parameters can be challenging. 
For example, the standard deviation on the fitted value of the allosteric
constant is rather large. On the other hand, this is mostly due to the low
sensitivity of the saturation function with respect to this parameter: even an estimation error of a factor two (which is still relatively small, and close to noise level) would shift the saturation curve by an amount that is comparable to the scattering in the available literature data, as shown in Figure~\ref{fig:l_sensitivity}.
\subsection*{Conclusions}
We have shown that a classic MWC allosteric model can successfully fit the 
saturation curves of calmodulin lobes and of the intact protein, and also 
account for the effects of several biologically relevant targets. 
This wide applicability was also achieved with a comparative economy of hypotheses
(independence of two domains, each in thermodynamic equilibrium between two 
conformations that have different affinities for the ligand and also different 
affinities for each target), thus providing a useful conceptual framework upon 
which further modelling work can be constructed.
A crucial point is that the apparent affinity of calmodulin is modulated by
simultaneous dynamic equilibria with different targets that exhibit a preference
(tighter binding) to one of the possible conformations of calmodulin.

\newpage
\section*{Methods}
\subsection*{MWC model of an isolated calmodulin lobe modulated by targets}
\begin{figure}[!h]
\begin{center}
\includegraphics[width=0.9\textwidth]{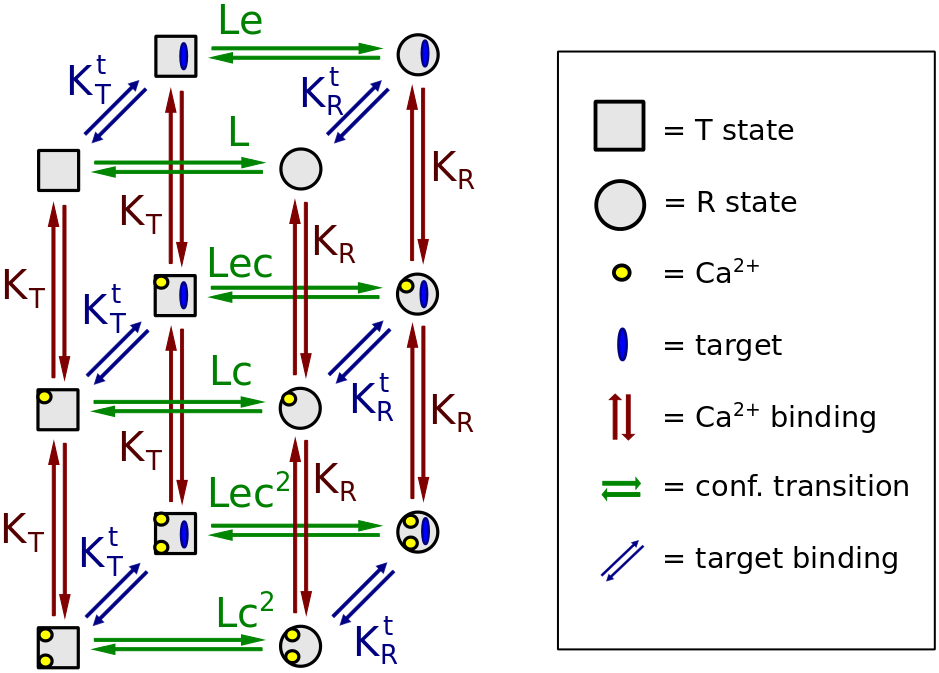}
\caption{
{\bf Diagram of the model used to represent isolated lobes of calmodulin (TR1C and TR2C)}. Each lobe has two binding sites for calcium, two possible conformations, and can bind targets with different affinities depending on its conformation. As a simplifying assumption, the two calcium-binding sites were assumed to have identical
affinities. The molecule can exist in two conformational states, T and R.
In the absence of ligand, the equilibrium constant for the R to T transition is the allosteric constant L. The R state has affinity $K_R$ for calcium and $K_R^t$ for target $t$. The T state has affinity $K_T$ for calcium and $K_T^t$ for target $t$. In accordance to classic linkage theory, each calcium ion bound to the molecule shifts the conformational equilibrium towards the state with the higher calcium affinity, causing a scaling of the allosteric constant L by a factor $c=K_R/K_T$. In an analogous fashion, the binding of a target shifts the conformational equilibrium 
towards the state that binds $t$ with the higher affinity, and scales
L by a factor $e=K_R^t / K_T^t$.
The model of intact calmodulin was obtained by combining two of these sub-models.}. \label{fig:model_diagram}
\end{center}
\end{figure}

Each isolated lobe of calmodulin exhibits cooperativity between its two calcium 
binding sites. We chose to model the isolated C-lobe (TR2C tryptic fragment) as
a two-subunit, two-state MWC molecule, where the subunits undergo concerted 
conformational transitions. 
As a starting point, we assumed that both calcium-binding sites on the
C-lobe were identical, as was observed for example for the CaM85/112
mutant \cite{Tan1996}. Therefore, in our model, both sites have calcium 
affinity $K_T$ when they are in the T state, and a higher affinity $K_R$ 
when the lobe is in the R state.
Targets can bind both conformations of the molecule, but with different 
affinities, as shown by several experimental studies \cite{Bayley1996,
Theoharis2008}.
A diagram of the resulting model is given in Figure~\ref{fig:model_diagram}.

\begin{figure}[!h]
\begin{center}
\includegraphics[width=0.8\textwidth]{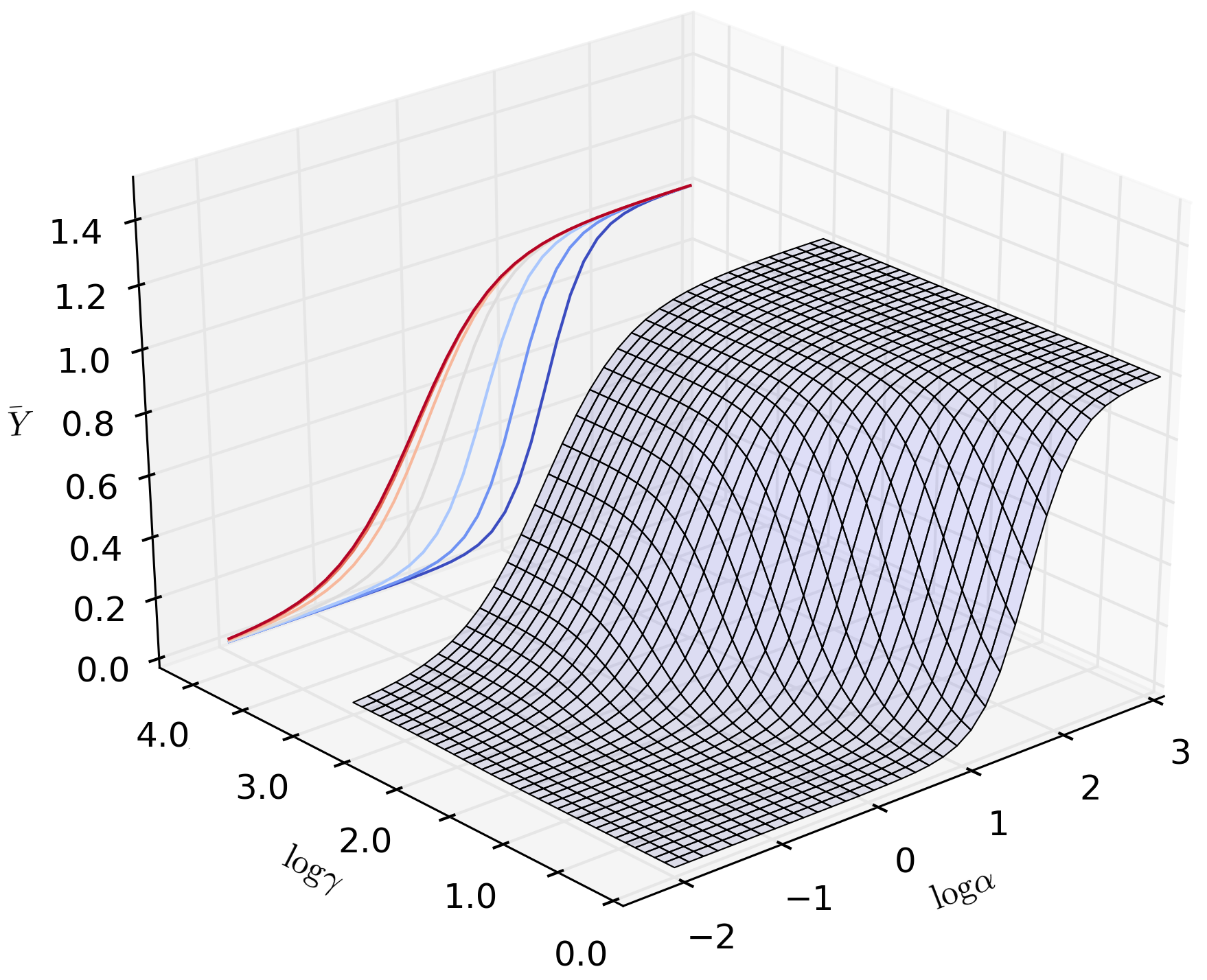}
\end{center}
\caption{
{\bf Saturation function of an MWC allosteric model} of a two-state molecule 
with 2 binding sites, such as TR2C, in the presence of a target with preference
for the R state. Increasing concentrations of target produce a left
shift in the calcium saturation curve.
}\label{fig:rubin_changeux_surf}
\end{figure}

Under the simplifying assumption, used throughout this work, that the two 
calcium-binding sites are identical, the following equations can be derived 
for the saturation function of TR2C in the presence of an allosteric target A
\cite{Rubin1966}:
\begin{align}\label{eq:rubin_changeux}
\setlength{\jot}{30pt} 
 \bar{Y} &= \frac{\alpha(1+\alpha)+L'\alpha c(1+\alpha c)}
                        {(1+\alpha)^2+L'(1+\alpha c)^2} \\
 L' &= L\left( \frac{1+\gamma e}{1 +\gamma} \right)^2                        
\end{align}
\noindent
where: $L=T_0/R_0$ is the allosteric constant in the absence of targets; $L'$ is
the allosteric constant in the presence of targets; $\alpha = [X]_{free}/K_R$ is
the ratio between the ligand affinity of the R state and the concentration of free
ligand; $c=K_R/K_T$ is the ration of the ligand affinities of the R and T state;
$\gamma=[A]_{free}/K^t_R$ is the ratio between the concentration of free
allosteric target and the target affinity of the R state; $e=K_R^t/K_T^t$ is 
the ratio between the target affinity of the R state and that of the T state.
The above equations allow the calculation of the saturation level of TR2C in the
presence of a known concentrations of \textit{free} ligand (calcium) and
allosteric targets (peptides). They also clearly show that the effect of targets
is equivalent to a modulation of the allosteric constant, whilst the other
parameters remain unaffected. 
The function $\bar{Y}$ describes a surface in the plane $(\alpha,\gamma)$,
as shown in Figure~\ref{fig:rubin_changeux_surf}.
The ligand saturation of an allosteric molecule, at equilibrium and at a given
concentration of free ligand and free target, is entirely determined
by the following parameters (note that all affinities are expressed as
\emph{dissociation} constants):
\begin{itemize}
 \item[-]the allosteric constant $L$ (or isomerisation constant), defined as
the concentration ratio between the R and T states in the absence of ligand.
 \item[-]the ligand affinity of the binding sites in the T state, $K_T$;
 \item[-]the affinity change upon transition from T to R state, $c=K_R/K_T$;
 \item[-]the affinity for the target when the molecule is in the T state,
$K^t_T$;
 \item[-]the change of affinity for the target upon transition from the T state
to  the R state, $e=K^t_R/K^t_T$.
\end{itemize}
A diagram of an MWC model of TR2C is given in Figure~\ref{fig:model_diagram}.
However, most available experimental data were obtained performing calcium
titrations of calmodulin the presence of a fixed \textit{total} concentration of
target, and therefore Equation~\ref{eq:rubin_changeux} was not directly applicable 
for fitting purposes.
This apparent difficulty can be circumvented analytically, by calculating 
the corresponding concentrations of free target, as shown for example by
Martinez et al. \cite{Martinez2000}.
However, we found convenient to follow a computational route, directly
implementing the model in the biochemical pathway simulator COPASI, which 
supports parameter fitting of steady-state properties when a system is
initialised with total concentrations of reagents in a reaction compartment.
The analytical model was instead used for a preliminary study on the general 
behavior of an MWC model for several choices of parameters, and also to
discriminate between alternative possible parameterisations, as shown in the 
Results section.

\subsection*{Choice of experimental data for parameter fitting and validation}
Several published steady-state titration curves described how
calmodulin and TR2C fractional saturation at varying concentrations of free
calcium change in the presence of different peptides that mimic the consensus 
motif of several \textit{in vivo} binding partners of calmodulin. 
Published datasets were digitised using the freely available PlotDigitizer
program. The datasets of Bayley et al. \cite{Bayley1996} and Shea et al.
\cite{Theoharis2008} were of particular interest because they reported
the affinity for the peptides both in the absence and at saturating
concentrations of calcium. In an allosteric perspective, these data provide 
estimates of the intrinsic affinity of the peptide for the T-state
(predominant in calcium-free calmodulin) and R-state (predominant in 
calcium-saturated calmodulin). These affinities were thus known quantities 
in our model.
The properties of the peptides used in the above mentioned experiments are 
summarised in Table~\ref{tab:peptides_summary}.
It must be noted that none of the experimental datasets in the literature 
reported error bars or standard deviations for the measured
datapoints, even when the datapoints were averages of measurements performed 
in triplicate \cite{Theoharis2008}. In reality, all plotted datapoints are
affected by uncertainty both on the x (free calcium concentration) and y axis
(saturation) although free calcium concentration is usually known to high
precision \cite{Peersen1997}.

\subsection*{Fitting procedure and choice of constraints on the parameter space}
When choosing the initial conditions for the fitting procedure we assumed that
our model must account for a wide range of observed behaviors.
An initial exploratory study on different possible parameterizations was
performed using the analytical MWC model described in the previous paragraphs,
using custom Python scripts. These preliminary results were used to formulate
additional constraints on the parametric space that the computational model was
allowed to sample fitted to the experimental data describing the saturation curve 
of TR2C in the presence of targets.
In the MWC framework, the apparent ligand affinity is produced by a mixture of 
two distinct populations of molecules, in the R-state and T-state. 
Target peptides shift the saturation curve by differentially stabilising the R
and T state. Therefore, any observed saturation curve, with or without targets, 
is bound to lie between two limit curves corresponding to  populations of 100\% 
R state, and 100\% T state. 
The two fully-stabilised states also exhibit no cooperativity, because no
further population shift is possible.
These observations provide a simple way to determine lower bounds for $K_T$ and
upper bounds for $K_R$, which in turn provide an upper bound for  $c = K_R/K_T$.
The saturation curve for TR2C in the presence of Nav1.2IQp showed an apparent
affinity of about $10\mu M$ and a  Hill coefficient greater than 1, indicating that the
T-state was not fully stabilised \cite{Theoharis2008}. 
Therefore the affinity of the pure T-state must be lower than $10\mu M$. 
On the other hand, data obtained with full-length calmodulin, showed that some 
peptides can increase affinity and decrease cooperativity, shifting the
saturation curve to the left to an extent that implies that the calcium affinity of the
pure R state must to be higher than 20nM \cite{Peersen1997}, assuming that the C-lobe
is responsible for the higher affinity when CaM is bound to a target, as shown by
data published by Shea et al. \cite{Evans2009}. 
Taken together, these observations led to the following constraints on the
allowed parameter values:
\begin{align}
K_R & \leq 20nM \\
K_T & \geq 10\mu M \\
c   &= \frac{K_R}{K_T} \leq 0.002.
\end{align}
\noindent

The chosen score function to minimise was the sum of squared residuals between
the predictions of the model and the measured values. The fitting was performed
using the built-in parameter estimation functions of COPASI, using 1000
iteration of a genetic algorithm with stochastic ranking \cite{Runarsson2000}, 
with a population size of 20, and with the constraints on the parameter values
described in the previous paragraph. A genetic algorithm is a non-local 
optimisation method and is therefore less prone to converging to local minima 
in comparision to gradient-based methods.

\subsection*{Calculation of Hill coefficients}
The Hill coefficients of calmodulin and calmodulin in the presence of targets 
were calculated as described in Ref.~\cite{Edelstein2010}, as the slope of the 
saturation function in the Hill plane:
\begin{equation}\label{eq:hill_coeff}
 n_H = \frac{d\left(\log \frac{\overline{Y}}{1 - \overline{Y}} \right)}
            {d(\log \alpha)}\phantom{\frac{1}{1}}
\end{equation}
\noindent
where $\alpha = [Ca^{2+}]_{free} / K_R$ is the free ligand concentration normalised by
the ligand affinity of the R state, and the fractional saturation function, 
$\overline{Y}$, is defined as the ration between the number of occupied binding
sites and the total number of binding sites, at a given concentration of free
ligand, and can be calculated as: 
\begin{equation}
 \overline{Y} = \frac{[Ca^{2+}]_{bound}}{2\cdot [TR2C]}
\end{equation}
\noindent
For an MWC model with two identical subunits, the Hill coefficient as a function of
ligand concentration describes a symmetric bell-shaped curve, with its maximum at
the point of half-saturation.
\subsection*{Parameterization of isolated N-lobe}
The model of isolated N-lobe (corresponding to the TR1C tryptic fragment) was formally identical to the  TR2C fragment described in the previous paragraphs. However, the parameter estimation required a different approach.
We could not directly fit the model to saturation curves in the absence and presence of targets, because the affinity of the isolated N-lobe for targets is very low \cite{Bayley1996}, and the resulting shift of the saturation curve very weak \cite{Theoharis2008}. Moreover, targets that bind calmodulin in calcium-free conditions have little or no interaction with the N-lobe, and their effect on its saturation curve was negligible.
First we assumed, for the sake of simplicity, that the two calcium-binding sites of the N-lobe are identical (as we did for the C-lobe). 
Based on the experimental evidence \cite{Peersen1997}, that the two lobes have very similar affinity when bound to high-affinity targets, we made the additional assumption  that, when in the R state, both lobes had the same calcium affinity. The affinity of the C-lobe in the R state was already known from the parameterization of the TR2C model.
We estimated the calcium affinity of the N-lobe in the T state by fitting the saturation curve by Grabarek and coworkers for the NCaM41/75 mutant (where the N-lobe was constrained in a closed conformation by a disulfide bond) with a non-cooperative model with two identical binding sites. 
Knowing the affinity of both the R and T state, the allosteric parameter $c = K_R / K_T$  was readily calculated. The only remaining free parameter in the model was the allosteric constant L, which was estimated by fitting and analytical MWC model onto the available experimental points for the saturation curve of the N-lobe \cite{Grabarek2005,VanScyoc2002,Evans2009}.
\subsection*{Model of intact calmodulin}
\subsubsection*{Generation of the model in SBML format}
The model of intact calmodulin was built by assuming that each lobe still behaved as it would in its isolated form.
When referring to the conformation of intact calmodulin, we adopt the following convention: "RT" means that the N lobe is in the R state, and the C-lobe is in the T state. Since the two lobes have different affinities and therefore saturate at different calcium concentrations, asymmetric states are biologically plausible. However, the resulting kinetic model is much more complex than the submodels comprising only one lobe.
Two largely independent lobes imply that the conformations are four: RR,RT,TR,TT. 
The binding sites are also 4 (A,B on the N lobe and C,D on the C-lobe), which leads to 64 possible states for calmodulin only (see Supplementary Material). With the addition of targets, combinatorial explosion implies that the number of equations in the kinetic model quickly increases to more than one thousand.
Manual modification of such models and inclusion of additional targets would be error-prone and labour-intensive. On the other hand, the interaction rules underlying the model are simple, and the complexity is purely combinatorial, which means the model lends itself to be generated iteratively.
Taking advantage of the existing  libSBML library, we wrote custom Python scripts to automatically generate allosteric models of calmodulin in the presence of targets, in the SBML format.
SBML is natively supported by COPASI, thus making the setup process of different kinetic models both faster and more reliable.

\subsubsection*{Affinity of targets for asymmetric conformations of calmodulin}
Our model allows for hemiconcerted conformational transitions (i.e. the two lobes can have different conformations, but the EF-hands on the same lobe are always in the same state). This poses no problem for calcium binding events, because calcium only binds to individual sites, whose affinity is determined by their state.
Targets, on the other hand, being bigger molecules (either peptides of whole proteins) bind to the whole molecule and usually interact with both lobes at once.
Their affinity for a given conformational state of calmodulin can be expected to be an interplay of their affinity for the two lobes.
A modelling challenge is posed by the fact that in the available experimental data, the affinity of calmodulin for a given target was usually measured only in two limit cases, in the absence of calcium (when almost all calmodulin will be in the TT state) and at saturating concentrations of calcium (when calmodulin will be mostly in the RR state).
In our model we also needed the affinities for the two asymmetric states RT and TR.
To overcome this limitations, we observed that targets that bind preferentially the calcium-saturated calmodulin exhibit a much stronger interaction with the C-lobe than with the N-lobe \cite{Drabikowski1982,Evans2009} and that the N-lobe alone can only bind very weakly to targets, and the affinity is in the millimolar range in the absence of calcium \cite{Bayley1996,Theoharis2008}. 
Collectively, these observations led to the following simplifying assumptions: 
\begin{itemize}
\item[-] The affinity of the target for the RR state is equal to the affinity for CaM, measured at saturating calcium concentrations;
\item[-] The affinity of the target for the TT state is equal to the affinity for CaM, measured in the absence of calcium. 
\item[-] The affinity of the target for the TR state is equal to the affinity for the TR2C fragment (isolated C lobe) at saturating calcium concentrations.
\item[-] The affinity of the target for the RT state is roughly equal to the affinity for the TT state
\end{itemize}
\noindent
With this set of assumptions, the model could reproduce several experimental saturation curves (see Results).

\section*{Acknowledgments}
ML was supported by the European Union Seventh Framework Programme SYNSYS (Synaptic Systems: dissecting brain function in health and disease) grant agreement 42167


\begin{thebibliography}{10}
\providecommand{\url}[1]{\texttt{#1}}
\providecommand{\urlprefix}{URL }
\expandafter\ifx\csname urlstyle\endcsname\relax
  \providecommand{\doi}[1]{doi:\discretionary{}{}{}#1}\else
  \providecommand{\doi}{doi:\discretionary{}{}{}\begingroup
  \urlstyle{rm}\Url}\fi
\providecommand{\bibAnnoteFile}[1]{%
  \IfFileExists{#1}{\begin{quotation}\noindent\textsc{Key:} #1\\
  \textsc{Annotation:}\ \input{#1}\end{quotation}}{}}
\providecommand{\bibAnnote}[2]{%
  \begin{quotation}\noindent\textsc{Key:} #1\\
  \textsc{Annotation:}\ #2\end{quotation}}
\providecommand{\eprint}[2][]{\url{#2}}

\bibitem{Faga2003}
Faga LA, Sorensen BR, VanScyoc WS, Shea MA (2003) {Basic interdomain boundary
  residues in calmodulin decrease calcium affinity of sites I and II by
  stabilizing helix-helix interactions.}
\newblock Proteins: Struct, Funct, Bioinf 50: 381--91.
\input{Faga2003}

\bibitem{Zhang1995}
Zhang M, Tanaka T, Ikura M (1995) {Calcium-induced conformational transition
  revealed by the solution structure of apo calmodulin}.
\newblock Nat Struct Mol Biol 2: 758--767.
\input{Zhang1995}

\bibitem{Grabarek2005}
Grabarek Z (2005) {Structure of a trapped intermediate of calmodulin: calcium
  regulation of EF-hand proteins from a new perspective.}
\newblock J Mol Biol 346: 1351--66.
\input{Grabarek2005}

\bibitem{Nelson1998}
Nelson MR, Chazin WJ (1998) {Conformational changes in Ca2+ sensor proteins}.
\newblock Protein Sci 7: 270--282.
\input{Nelson1998}

\bibitem{Rhoads1997}
Rhoads R (1997) {Sequence motifs for calmodulin recognition}.
\newblock FASEB J 11: 331--340.
\input{Rhoads1997}

\bibitem{Peersen1997}
Peersen OB, Madsen TS, Falke JJ (1997) {Intermolecular tuning of calmodulin by
  target peptides and proteins: differential effects on Ca2+ binding and
  implications for kinase activation.}
\newblock Protein Sci 6: 794--807.
\input{Peersen1997}

\bibitem{Bayley1996}
Bayley PM, Findlay WA, Martin SR (1996) {Target recognition by calmodulin:
  dissecting the kinetics and affinity of interaction using short peptide
  sequences.}
\newblock Protein Sci 5: 1215--28.
\input{Bayley1996}

\bibitem{Gaertner2004}
Gaertner TR, A PJ, Waxham MN (2004) {RC3/Neurogranin and
  Ca2+/calmodulin-dependent protein kinase II produce opposing effects on the
  affinity of calmodulin for calcium.}
\newblock J Biol Chem 279: 39374--82.
\input{Gaertner2004}

\bibitem{Evans2009}
Evans TIA, Shea MA (2009) {Energetics of Calmoudlin Domain Interactions with
  the Calmodulin Binding Domain of CaMKII}.
\newblock Proteins: Struct, Funct, Bioinf 76: 47--61.
\input{Evans2009}

\bibitem{Feldkamp2011}
Feldkamp MD, Yu L, Shea MA (2011) {Structural and Energetic Determinants of Apo
  Calmodulin Binding to the IQ Motif of the NaV 1.2 Voltage-Dependent Sodium
  Channel}.
\newblock Structure 19: 733--747.
\input{Feldkamp2011}

\bibitem{Kim2005}
Kim SA, Heinze KG, Bacia K, Waxham MN, Schwille P (2005) {Two-photon
  cross-correlation analysis of intracellular reactions with variable
  stoichiometry.}
\newblock Biophys J 88: 4319--36.
\input{Kim2005}

\bibitem{Theoharis2008}
Theoharis NT, Sorensen BR, Theisen-Toupal J, Shea MA (2008) {The neuronal
  voltage-dependent sodium channel type II IQ motif lowers the calcium affinity
  of the C-domain of calmodulin.}
\newblock Biochemistry 47: 112--23.
\input{Theoharis2008}

\bibitem{O'Donnell2011}
O'Donnell SE, Yu L, Fowler CA, Shea MA (2011) {Recognition of
  $\beta$-calcineurin by the domains of calmodulin: thermodynamic and
  structural evidence for distinct roles.}
\newblock Proteins: Struct, Funct, Bioinf 79: 765--86.
\input{O'Donnell2011}

\bibitem{Monod1965}
Monod JL, Wyman J, Changeux JP (1965) {On the Nature of Allosteric Transitions
  : A Plausible Model}.
\newblock J Mol Biol 12: 88--118.
\input{Monod1965}

\bibitem{Rubin1966}
Rubin MM, Changeux JP (1966) {On the nature of allosteric transitions:
  implications of non-exclusive ligand binding.}
\newblock J Mol Biol 21: 265--74.
\input{Rubin1966}

\bibitem{Malmendal1999}
Malmendal A, Even\"{a}s J, Fors\'{e}n S, Akke M (1999) {Structural dynamics in
  the C-terminal domain of calmodulin at low calcium levels.}
\newblock J Mol Biol 293: 883--99.
\input{Malmendal1999}

\bibitem{Evenas1999}
Even\"{a}s J, Fors\'{e}n S, Malmendal A, Akke M (1999) {Backbone dynamics and
  energetics of a calmodulin domain mutant exchanging between closed and open
  conformations.}
\newblock J Mol Biol 289: 603--17.
\input{Evenas1999}

\bibitem{Evenas2001}
Even\"{a}s J, Malmendal A, Akke M (2001) {Dynamics of the transition between
  open and closed conformations in a calmodulin C-terminal domain mutant.}
\newblock Structure 9: 185--95.
\input{Evenas2001}

\bibitem{Chen2007}
Chen YG, Hummer G (2007) {Slow conformational dynamics and unfolding of the
  calmodulin C-terminal domain.}
\newblock J Am Chem Soc 129: 2414--5.
\input{Chen2007}

\bibitem{Tripathi2009}
Tripathi S, Portman JJ (2009) {Inherent flexibility determines the transition
  mechanisms of the EF-hands of calmodulin}.
\newblock Proc Natl Acad Sci USA 106: 2104--2109.
\input{Tripathi2009}

\bibitem{Fallon2003}
Fallon J, Quiocho F (2003) {A Closed Compact Structure of Native
  Ca2+-Calmodulin}.
\newblock Structure 11: 1303--1307.
\input{Fallon2003}

\bibitem{Tan1996}
Tan RY, Mabuchi Y, Grabarek Z (1996) {Blocking the Ca2+-induced conformational
  transitions in calmodulin with disulfide bonds.}
\newblock J Biol Chem 271: 7479--83.
\input{Tan1996}

\bibitem{Meyer1996}
Meyer DF, Mabuchi Y, Grabarek Z (1996) {The Role of Phe-92 in the Ca2+ -induced
  Conformational Transition in the C-terminal Domain of Calmodulin}.
\newblock Biochemistry 271: 11284--11290.
\input{Meyer1996}

\bibitem{Ababou2001}
Ababou A, Desjarlais JR (1996) {Solvation energetics and conformational change
  in EF-hand proteins}.
\newblock Protein Sci 2: 301--312.
\input{Ababou2001}

\bibitem{Ikura2006}
Ikura M, Ames JB (2006) {Genetic polymorphism and protein conformational
  plasticity in the calmodulin superfamily: two ways to promote
  multifunctionality.}
\newblock Proc Natl Acad Sci USA 103: 1159--64.
\input{Ikura2006}

\bibitem{Kumar2013b}
Kumar V, Chichili VPR, Zhong L, Tang X, Velazquez-Campoy A, et~al. (2013)
  {Structural basis for the interaction of unstructured neuron specific
  substrates neuromodulin and neurogranin with Calmodulin.}
\newblock Sci Rep 3: 1392.
\input{Kumar2013b}

\bibitem{Ikura1992}
Ikura M, Clore GM, Gronenborn AM, Zhu G, Klee CB, et~al. (1992) {Solution
  structure of a calmodulin-target peptide complex by multidimensional NMR}.
\newblock Science 256: 632--8.
\input{Ikura1992}

\bibitem{Schumacher2001}
Schumacher M, Rivard AF, Bachinger HP, Adelman JP (2001) {Structure of the
  gating domain of a Ca2+-activated K+ channel complexed with Ca2+/calmodulin.}
\newblock Nature 410: 1120-1124.
\input{Schumacher2001}

\bibitem{VanDerSpoel1996}
Van~der Spoel D, De~Groot BL, Hayward S, Berendsen HJC, Vogel H (1996) {Bending
  of the Calmodulin central helix: A theoretical study}.
\newblock Protein Sci 5: 2044--2053.
\input{VanDerSpoel1996}

\bibitem{Stefan2008}
Stefan MI, Edelstein SJ, {Le Nov\`ere} N (2008) {An allosteric model of
  calmodulin explains differential activation of PP2B and CaMKII}.
\newblock PNAS 31.
\input{Stefan2008}

\bibitem{Finn1995}
Finn BE, Evenas J, Drakenberg T, Waltho JP, Thulin E, et~al. (1995)
  {Calcium-induced structural changes and domain authonomy in calmodulin}.
\newblock Nat Struct Biol 2: 777--783.
\input{Finn1995}

\bibitem{Minowa1984}
Minowa O, Yagi K (1984) {Calcium binding to tryptic fragments of calmodulin.}
\newblock J Biochem 96: 1175--82.
\input{Minowa1984}

\bibitem{Linse1991}
Linse S, Helmersson A, Fors\'{e}n S (1991) {Calcium binding to calmodulin and
  its globular domains.}
\newblock J Biol Chem 266: 8050--4.
\input{Linse1991}

\bibitem{Nelson2002}
Nelson MR, Thulin E, Fagan PA, Fors\'{e}n S, Chazin WJ (2002) {The EF-hand
  domain : A globally cooperative structural unit}.
\newblock Protein Sci 11: 198--205.
\input{Nelson2002}

\bibitem{VanScyoc2002}
VanScyoc WS, Sorensen BR, Rusinova E, Laws WR, Ross JB, et~al. (2002) {Calcium
  binding to calmodulin mutants monitored by domain-specific intrinsic
  phenylalanine and tyrosine fluorescence.}
\newblock Biophys J 83: 2767--80.
\input{VanScyoc2002}

\bibitem{Faas2011}
Faas GC, Raghavachari S, Lisman JE, Mody I (2011) {Calmodulin as a direct
  detector of Ca2+ signals.}
\newblock Nat Neurosci 14: 301--4.
\input{Faas2011}

\bibitem{Jiang2010}
Jiang J, Zhou Y, Zou J, Chen Y, Patel P, et~al. (2010) {Site-specific
  modification of calmodulin Ca²(+) affinity tunes the skeletal muscle
  ryanodine receptor activation profile.}
\newblock Biochem J 432: 89--99.
\input{Jiang2010}

\bibitem{Chen2011}
Chen Y, Zhou Y, Lin X, Wong HC, Xu Q, et~al. (2011) {Molecular interaction and
  functional regulation of connexin50 gap junctions by calmodulin.}
\newblock Biochem J 435: 711--22.
\input{Chen2011}

\bibitem{Zhou2010}
Zhou Y, Tzeng WP, Wong HC, Ye Y, Jiang J, et~al. (2010) {Calcium-dependent
  association of calmodulin with the rubella virus nonstructural protease
  domain.}
\newblock J Biol Chem 285: 8855--68.
\input{Zhou2010}

\bibitem{Drake1996}
Drake SK, Falke JJ (1996) {Kinetic Tuning of the EF-Hand Calcium Binding Motif
  : The Gateway Residue Independently Adjusts (i) Barrier Height and (ii)
  Equilibrium}.
\newblock Biochemistry 35: 1753--1760.
\input{Drake1996}

\bibitem{Drake1997}
Drake SK, Zimmer MA, Kundrot C, Falke JJ (1997) {Molecular tuning of an
  EF-hand-like calcium binding loop. Contributions of the coordinating side
  chain at loop position 3.}
\newblock J Gen Physiol 110: 173--84.
\input{Drake1997}

\bibitem{Wang2004}
Wang BO, Martin SR, Newman RA, Hamilton SL, Shea MA, et~al. (2004) {Biochemical
  properties of V91G calmodulin : A calmodulin point mutation that deregulates
  muscle contraction in Drosophila}.
\newblock Protein Sci 13: 3285--3297.
\input{Wang2004}

\bibitem{Kubota2007}
Kubota Y, Putkey Ja, Waxham MN (2007) {Neurogranin controls the spatiotemporal
  pattern of postsynaptic Ca2+/CaM signaling.}
\newblock Biophys J 93: 3848--59.
\input{Kubota2007}

\bibitem{Cui2003}
Cui Y, Wen J, Sze KH, Man D, Lin D, et~al. (2003) {Interaction between
  calcium-free calmodulin and IQ motif of neurogranin studied by nuclear
  magnetic resonance spectroscopy}.
\newblock Anal Biochem 315: 175 - 182.
\input{Cui2003}

\bibitem{Chichili2013}
Chichili VPR, Xiao Y, Seetharaman J, Cummins TR, Sivaraman J (2013) {Structural
  Basis for the Modulation of the Neuronal Voltage-Gated Sodium Channel NaV1.6
  by Calmodulin }.
\newblock Sci Rep 3.
\input{Chichili2013}

\bibitem{Stefan2009}
Stefan MI, Edelstein SJ, {Le Nov\`ere} N (2009) {Computing phenomenologic
  Adair-Klotz constants from microscopic MWC parameters}.
\newblock BMC Syst Biol 68.
\input{Stefan2009}

\bibitem{BioModels2010}
Li C, Donizelli M, Rodriguez N, Dharuri H, Endler L, et~al. (2010) {BioModels
  Database: An enhanced, curated and annotated resource for published
  quantitative kinetic models.}
\newblock BMC Sys Bio 4: 92.
\input{BioModels2010}

\bibitem{Johnson1996}
Johnson JD, Snyder C, Walsh M, Flynn M (1996) {Effects of myosin light chain
  kinase and peptides on Ca2+ exchange with the N- and C-terminal Ca2+ binding
  sites of calmodulin.}
\newblock J Biol Chem 271: 761--7.
\input{Johnson1996}

\bibitem{Drabikowski1982}
Drabikowski W, Brzeska H, Venyaminov SY (1982) {Tryptic Fragments of
  Calmodulin}.
\newblock J Biol Chem 257: 11584--11590.
\input{Drabikowski1982}

\bibitem{Porumb1994}
Porumb T, Yau P, Harvey TS, Ikura M (1994) {A calmodulin-target peptide hybrid
  molecule with unique calcium-binding properties.}
\newblock Protein Eng 7: 109--15.
\input{Porumb1994}

\bibitem{Ulmer2003}
Ulmer TS, Soelaiman S, Li S, Klee CB, Tang WJ, et~al. (2003) {Calcium
  dependence of the interaction between calmodulin and anthrax edema factor.}
\newblock J Biol Chem 278: 29261--6.
\input{Ulmer2003}

\bibitem{Persechini2002}
Persechini A, Stemmer PM (2002) {Calmodulin is a limiting factor in the cell.}
\newblock Trends Cardiovasc Med 12: 32--7.
\input{Persechini2002}

\bibitem{Kumar2013a}
Kumar V, Chichili VPR, Tang X, Sivaraman J (2013) A novel trans conformation of
  ligand-free calmodulin.
\newblock PLoS One 8: e54834.
\input{Kumar2013a}

\bibitem{Putkey2003}
Putkey JA, Kleerekoper Q, Gaertner TR, Waxham MN (2003) {A new role for IQ
  motif proteins in regulating calmodulin function.}
\newblock J Biol Chem 278: 49667--70.
\input{Putkey2003}

\bibitem{Pedigo1995}
Pedigo S, Shea MA (1995) {Discontinuous equilibrium titrations of cooperative
  calcium binding to calmodulin monitored by 1-D 1H-nuclear magnetic resonance
  spectroscopy.}
\newblock Biochemistry 34: 10676--89.
\input{Pedigo1995}

\bibitem{Martinez2000}
Martinez KL, Corringer PJ, Edelstein SJ, Changeux JP, M\'{e}rola F (2000)
  {Structural differences in the two agonist binding sites of the Torpedo
  nicotinic acetylcholine receptor revealed by time-resolved fluorescence
  spectroscopy.}
\newblock Biochemistry 39: 6979--90.
\input{Martinez2000}

\bibitem{Runarsson2000}
Runarsson TP (2000) {Stochastic ranking for constrained evolutionary
  optimization}.
\newblock IEEE Trans Evol Comput 4: 284--294.
\input{Runarsson2000}

\bibitem{Edelstein2010}
Edelstein SJ, Stefan MI, {Le Nov\`{e}re} N (2010) {Ligand depletion in vivo
  modulates the dynamic range and cooperativity of signal transduction.}
\newblock PLoS One 5: e8449.
\input{Edelstein2010}

\end{thebibliography}

\newpage


\section*{Supplementary material}

\subsection*{TR2C model validation}

\begin{figure}[!h]
\begin{center}
\includegraphics[width=0.65\textwidth]{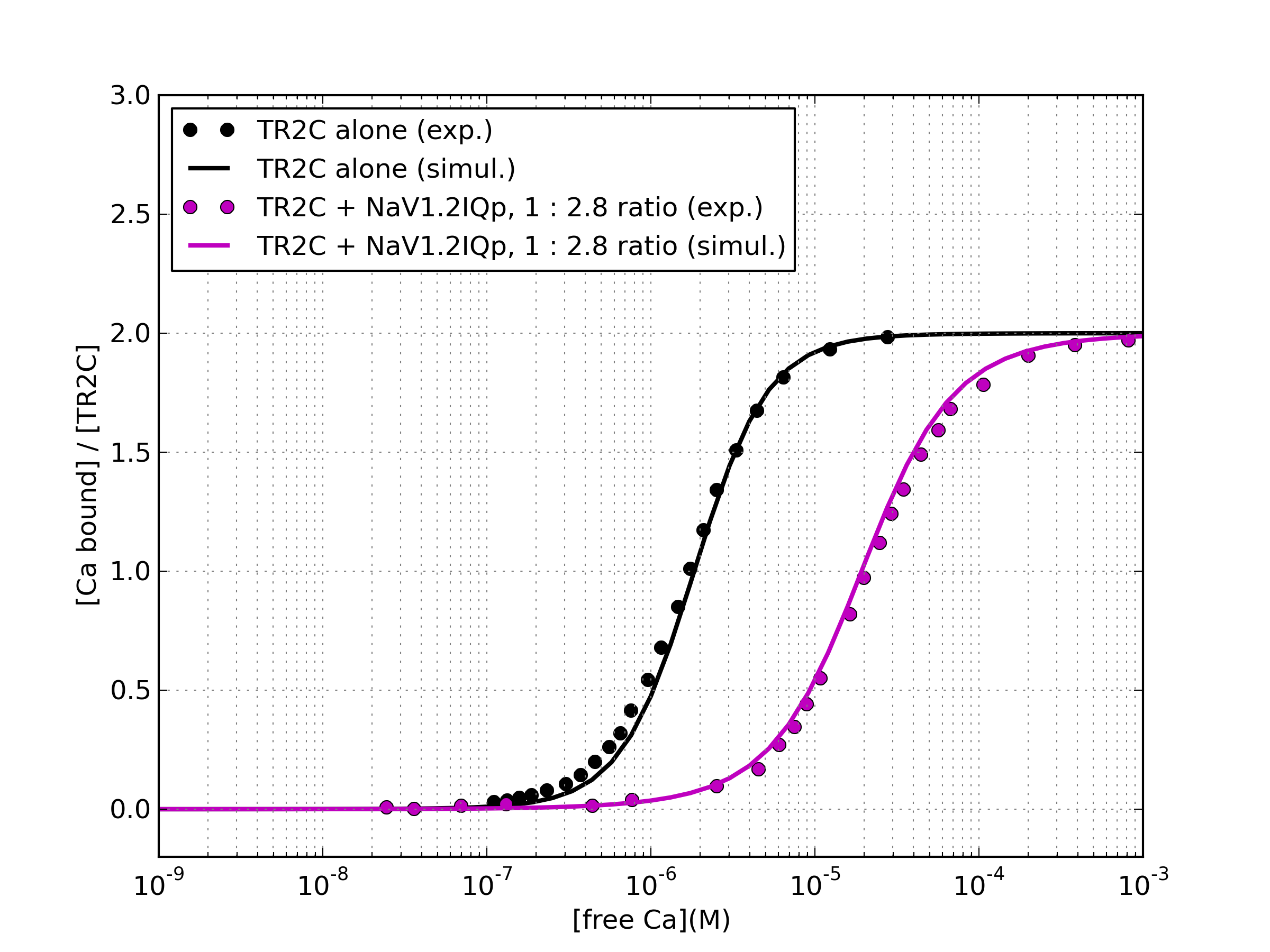}
\end{center}
\caption{
{\bf Prediction of the behavior of TR2C in the presence of the NaV1.2IQp peptide}, 
and comparison with experimental data not included in the original fitting procedure 
\cite{Theoharis2008}.}\label{fig:validation1}
\end{figure}

\begin{figure}[!h]
\begin{center}
\includegraphics[width=0.65\textwidth]{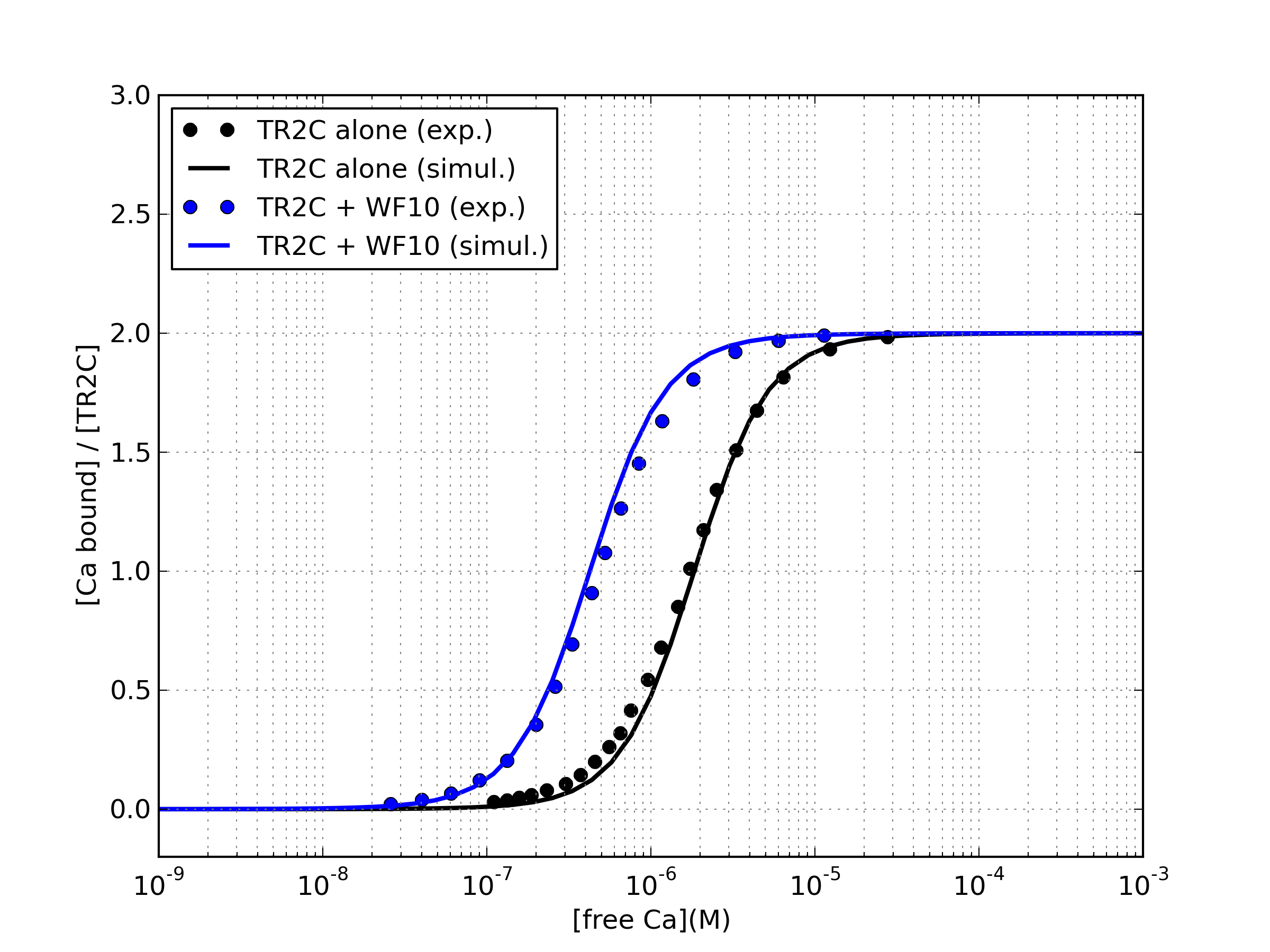}
\end{center}
\caption{
{\bf Prediction of the behavior of TR2C in the presence of the WFF peptide}, 
and comparison with experimental data not included in the original fitting procedure \cite{Bayley1996}.}\label{fig:validation2}
\end{figure}

\subsection*{Parameters for the TR2C model}
The Rubin-Changeux equations can be derived by recursively applying mass action
law to the equilibrium reactions, thus
calculating equilibrium concentrations for all possible conformations and
liganded states, as a function of the concentrations of free calcium and free
target. Kinetic parameters were necessary for implementing the model in COPASI,
were all reversible reactions must be split into the corresponding forward and
backward rate law, but their exact values were not relevant for the purpose of
this paper, where we only examined steady-state properties.
The model was thus written using arbitrary kinetic constants, chosen to ensure
that the ratio between the forward and backward rate constants of each
reversible reaction was equal to the equilibrium constant calculated by
recursive application of mass action law. The model for TR1C is formally
identical to that of TR2C and is omitted.

\begin{table}[h]
\begin{center}
\begin{tabular}{l l l l}
\hline
& & &  \\
Parameter & Description & Value & Source \\
\hline
& & &  \\
L 	& Allosteric constant & $8616.61$ & This paper \\
c 	& ratio of affinities & $0.000317019$ & This paper \\
lC 	& Allosteric constant when C site is occupied &2.73163 & Calculated as
lC = L*c\\
lD 	& Allosteric constant when D site is occupied &2.73163 & Calculated as
lC = L*c\\
lCD 	& Allosteric constant when C and D site are occupied & 0.000865977 &
Calculated as lC = L*c*c\\

KCR 	& calcium affinity of site C in the R state & $1.97874e-08$ & This paper
\\
KDR 	& calcium affinity of site D in the R state & $1.97874e-08$ & This paper
\\
KCT 	& calcium affinity of site C in the T state & $6.24171e-05$ & This paper
\\		
KDT 	& calcium affinity of site C in the T state & $6.24171e-05$ & This paper
\\
KdpepR 	& peptide affinity for the R state & $7.6e-08$ &
\cite{Bayley1996,Theoharis2008}\\
KdpepT 	& peptide affinity for the R state & $1.1e-05$ &
\cite{Bayley1996,Theoharis2008}, this paper\\
& & &  \\
\hline
\end{tabular}
\end{center}
\caption{Summary of the model parameters. The kinetic rates for the
conformational transitions were
calculated based on the value of the allosteric constant estimated in this work,
and the value of the exchange 
constant for the conformational fluctuations $k_{ex}\sim 20000s^{-1}$, estimated
by Malmendal and coworkers for the 
C-lobe \cite{Malmendal1999}. The exact value does not affect steady state
properties, but a solid estimate will be 
importat for future work where kinetic properties will also taken into account.}
\label{tab:param_summary}
\end{table}

\begin{table}[h]
\begin{center}
\begin{tabular}{l l}
\hline
&  \\
Species name & Description  \\
\hline
&  \\
ca	& calcium ion \\
pep	& generic name for target peptides (e.g. WFF, WF10, Nav1.2IQp)\\
R\_0	& calmodulin in the R state with no calcium bound\\
R\_C 	& calmodulin in the R state with calcium bound to site C\\
R\_D 	& calmodulin in the R state with calcium bound to site D\\
R\_CD 	& calmodulin in the R state with calcium bound to site C and D\\
T\_0	& calmodulin in the T state with no calcium bound\\
T\_C	& calmodulin in the T state with calcium bound to site C\\
T\_D 	& calmodulin in the T state with calcium bound to site D\\
T\_CD	& calmodulin in the T state with calcium bound to site C and D\\
pep\_R\_0	& peptide-bound calmodulin in the R state with no calcium
bound\\
pep\_R\_C 	& peptide-bound calmodulin in the R state with calcium bound to
site C\\
pep\_R\_D 	& peptide-bound calmodulin in the R state with calcium bound to
site D\\
pep\_R\_CD 	& peptide-bound calmodulin in the R state with calcium bound to
site C and D\\
pep\_T\_0	& peptide-bound calmodulin in the T state with no calcium
bound\\
pep\_T\_C	& peptide-bound calmodulin in the T state with calcium bound to
site C\\
pep\_T\_D	& peptide-bound calmodulin in the T state with calcium bound to
site D\\
pep\_T\_CD	& peptide-bound calmodulin in the T state with calcium bound to
sites C and D\\
&  \\
\hline
\end{tabular}
\end{center}
\caption{Summary of the chemical species in the model.}
\label{tab:species_summary}
\end{table}

\begin{table}[h]
\begin{center}
\begin{tabular}{l l}
\hline
&  \\
Reaction name & Description  \\
\hline
&  \\
pep binding to R\_0			& pep + R\_0 $\rightleftarrows$
pep\_R\_0 \\ 
pep binding to R\_C			& pep + R\_C $\rightleftarrows$
pep\_R\_C \\
pep binding to R\_D			& pep + R\_D $\rightleftarrows$
pep\_R\_D \\
pep binding to R\_CD			& pep + R\_{CD} $\rightleftarrows$
pep\_R\_{CD} \\
pep binding to T\_0			& pep + T\_0 $\rightleftarrows$
pep\_T\_0 \\
pep binding to T\_C			& pep + T\_C $\rightleftarrows$
pep\_T\_C \\
pep binding to T\_D			& pep + T\_D $\rightleftarrows$
pep\_T\_D \\
pep binding to T\_CD			& pep + T\_{CD} $\rightleftarrows$
pep\_T\_{CD} \\
ca binding to R\_0 on site C 		& ca + R\_0 $\rightleftarrows$ R\_C  \\
ca binding to R\_0 on site D 		& ca + R\_0 $\rightleftarrows$ R\_D  \\
ca binding to R\_C on site D 		& ca + R\_C $\rightleftarrows$ R\_CD \\
ca binding to R\_D on site C 		& ca + R\_D $\rightleftarrows$ R\_CD \\
ca binding to T\_0 on site C 		& ca + T\_0 $\rightleftarrows$ T\_C  \\
ca binding to T\_0 on site D 		& ca + T\_0 $\rightleftarrows$ T\_D  \\	
ca binding to T\_C on site D 		& ca + T\_C $\rightleftarrows$ T\_CD \\
ca binding to T\_D on site C 		& ca + T\_D $\rightleftarrows$ T\_CD \\
ca binding to pep\_R\_0 on site C 	& ca + pep\_R\_0 $\rightleftarrows$
pep\_R\_C   \\
ca binding to pep\_R\_0 on site D 	& ca + pep\_R\_0 $\rightleftarrows$
pep\_R\_D   \\
ca binding to pep\_R\_C on site D 	& ca + pep\_R\_C $\rightleftarrows$
pep\_R\_CD  \\
ca binding to pep\_R\_D on site C 	& ca + pep\_R\_D $\rightleftarrows$
pep\_R\_CD  \\
ca binding to pep\_T\_0 on site C 	& ca + pep\_T\_0 $\rightleftarrows$
pep\_T\_C   \\
ca binding to pep\_T\_0 on site D 	& ca + pep\_T\_0 $\rightleftarrows$
pep\_T\_D   \\
ca binding to pep\_T\_C on site D 	& ca + pep\_T\_C $\rightleftarrows$
pep\_T\_CD  \\
ca binding to pep\_T\_D on site C 	& ca + pep\_T\_D $\rightleftarrows$
pep\_T\_CD  \\
Conformational transition T\_0 / R\_0			& T\_0
$\rightleftarrows$ R\_0 \\
Conformational transition T\_C / R\_C			& T\_0
$\rightleftarrows$ R\_0  \\
Conformational transition T\_D / R\_D			& T\_0
$\rightleftarrows$ R\_0  \\
Conformational transition T\_CD / R\_CD		& T\_0 $\rightleftarrows$ R\_0 
\\
Conformational transition pep\_T\_0 / pep\_R\_0 	& pep\_T\_0
$\rightleftarrows$ pep\_R\_0  \\
Conformational transition pep\_T\_C / pep\_R\_C 	& pep\_T\_C
$\rightleftarrows$ pep\_R\_C  \\
Conformational transition pep\_T\_D / pep\_R\_D 	& pep\_T\_D
$\rightleftarrows$ pep\_ R\_D  \\
Conformational transition pep\_T\_CD / pep\_R\_CD 	& pep\_T\_CD
$\rightleftarrows$ pep\_ R\_CD  \\
&  \\
\hline
\end{tabular}
\end{center}
\caption{Summary of the reactions used in the model. All listed reactions are
reversible.}
\label{tab:reactions_summary}
\end{table}

\subsection*{Parameters for the model of intact calmodulin}

The generic names ``tbp'' and ``rbp'' were used to desctibe peptides that
exhibit a tighter binding for the T and R state, respectively.

The reactions of the model of intact calmodulin contains over 500 reversible
reactions (or more if additional targets are included), therefore they
are not listed here for brevity. The SBML model can however be retrived from 
Biomodels Database (model ID: MODEL1405060000).

\begin{table}[h]
\begin{center}
\begin{tabular}{l l l l}
\hline
& & &  \\
Parameter & Description & Value & Source\\
\hline
lC   &  Allosteric constant for the C-lobe  &	8616.61	& this paper 	\\
lN   &  Allosteric constant for the N-lobe  &	398000	& this paper 	\\
cC   &  defined as KCR / KCT  &	0.000159	& this paper 	\\
cN   &  defined as KAR / KAT  &	0.000215	& this paper 	\\

KAR & Ca affinity of site A in the R state & 1.97628e-08	& Calculated  as KAT*cN	 \\ 
KAT & Ca affinity of site A in the T state & 9.192e-05	   	& this paper \\ 	
KBR & Ca affinity of site B in the R state & 1.97628e-08	& Calculated  as KBT*cN \\
KBT & Ca affinity of site B in the T state & 9.192e-05		& this paper \\
KCR & Ca affinity of site C in the R state & 1.98496e-08	& Calculated  as KCT*cC \\
KCT & Ca affinity of site C in the T state & 0.00012484	& this paper 	\\
KDR & Ca affinity of site D in the R state & 1.98496e-08 	& Calculated  as KDT*cC \\
KDT & Ca affinity of site C in the T state & 0.00012484  	& this paper \\

eCR\_tbp & Kd\_tbp\_RR / Kd\_tbp\_RT &	(target-dependent)	& /     \\
eCT\_tbp	& Kd\_tbp\_TR / Kd\_tbp\_TT &	(target-dependent)	& / 	\\
eNR\_tbp	& Kd\_tbp\_RR / Kd\_tbp\_TR &	(target-dependent)	& / 	\\ 	
eNT\_tbp	& Kd\_tbp\_RT / Kd\_tbp\_TT &	(target-dependent)	& / 	\\ 

eCR\_rbp & Kd\_tbp\_RR / Kd\_tbp\_RT &	(target-dependent)	& / 	\\
eCT\_rbp	& Kd\_tbp\_TR / Kd\_tbp\_TT &	(target-dependent)	& /	\\
eNR\_rbp	& Kd\_tbp\_RR / Kd\_tbp\_TR &	(target-dependent)	& / 	\\ 	
eNT\_rbp	& Kd\_tbp\_RT / Kd\_tbp\_TT &	(target-dependent)	& /	\\    
\hline
\end{tabular}
\end{center}
\caption{Summary of the independent parameters for the model of intact
calmodulin. The kinetic rates for the conformational transitions were
calculated based on the value of the allosteric constant estimated in this work,
and the value of the exchange 
constant for the conformational fluctuations $k_{ex}\sim 20000s^{-1}$, estimated
by Malmendal and coworkers for the 
C-lobe \cite{Malmendal1999}. The exact value does not affect steady state
properties. The dummy names ``tbp'' and 
``rbp`` indicate the generic targets that bind preferably to the T and R states,
respectively.}
\label{tab:param_summary_intact}
\end{table}


\end{document}